\documentclass{ieeeaccess}
\usepackage{threeparttable}

\usepackage{amsmath,amssymb,amsfonts}
\usepackage{graphicx}
\usepackage{textcomp}
\usepackage{hyperref}
\usepackage{multirow}
\usepackage{array} 
\usepackage{booktabs} 
\hypersetup{hidelinks,pdftitle={Ensembling LLMs for AI-Augmented Cybersecurity Software Requirements Generation},pdfauthor={Santiago Perez-Acuna, Yod-Samuel Martin, Juan C. Yelmo}}

\usepackage{bm}
\makeatletter
\AtBeginDocument{\DeclareMathVersion{bold}
\SetSymbolFont{operators}{bold}{T1}{times}{b}{n}
\SetSymbolFont{NewLetters}{bold}{T1}{times}{b}{it}
\SetMathAlphabet{\mathrm}{bold}{T1}{times}{b}{n}
\SetMathAlphabet{\mathit}{bold}{T1}{times}{b}{it}
\SetMathAlphabet{\mathbf}{bold}{T1}{times}{b}{n}
\SetMathAlphabet{\mathtt}{bold}{OT1}{pcr}{b}{n}
\SetSymbolFont{symbols}{bold}{OMS}{cmsy}{b}{n}
\renewcommand\boldmath{\@nomath\boldmath\mathversion{bold}}}
\makeatother

\newsavebox{\fullDataSummaryBox}
\newsavebox{\oobPairedDeltasBox}

\newcommand{\diaghead}[1]{%
  \rotatebox[origin=l]{45}{\strut #1}%
}

\newcommand{\headk}{\texorpdfstring{\(k\)}{k}}

\vol{14}
\year{2026}

\def\BibTeX{{\rm B\kern-.05em{\sc i\kern-.025em b}\kern-.08em
    T\kern-.1667em\lower.7ex\hbox{E}\kern-.125emX}}
\begin{document}
\history{Date of publication xxxx 00, 0000, date of current version xxxx 00, 0000.}
\doi{10.1109/ACCESS.2026.Doi Number}

\title{Ensembling LLMs for AI-Augmented Cybersecurity Software Requirements Generation}

\author{\uppercase{Santiago Perez-Acuna}\authorrefmark{1}, \uppercase{Yod-Samuel Mart\'in}\authorrefmark{1}, \IEEEmembership{Member, IEEE}, AND \uppercase{Juan C. Yelmo}\authorrefmark{1}}

\address[1]{Information Processing and Telecommunications Center, ETSI Telecomunicación, Universidad Politécnica de Madrid, 28040 Madrid, Spain}
\tfootnote{This work was supported in part by the Spanish Ministry of Science and Innovation through the project Re-InITS (Sistemas informáticos industriales confiables en la era de la computación en el continuo) under Grant PID2024-155230OB-C43.}

\markboth
{Perez-Acuna \headeretal: Ensembling LLMs for AI-Augmented Cybersecurity Software Requirements Generation}
{Perez-Acuna \headeretal: Ensembling LLMs for AI-Augmented Cybersecurity Software Requirements Generation}

\corresp{Corresponding author: Yod-Samuel Martín (e-mail: ys.martin@upm.es).}

\begin{abstract}
Translating high-level controls from security standards into concrete, system-specific requirements is central to cybersecurity requirements engineering.
Large language models (LLMs) can accelerate this labor-intensive, recall-sensitive task, but any single run is unreliable: it misses valid safeguards while introducing plausible hallucinations, and outputs shift across runs and models.
We reframe this variability as a resource: rather than selecting one output, we study post-generation ensembling, aggregating stochastic runs with information-retrieval data-fusion operators.
We propose two strategies: Uniform fusion rewards mere cross-run agreement, whereas Naive-Bayes fusion weights each run by its estimated reliability.
We evaluate both over 24 runs from 12 configurations across four model families, generated for ten ISO/IEC~27002:2022 controls and expert-judged against a gold standard of 72 valid requirements.
Pooling every run's output recovers all 72---whereas single configurations recover on average under half---but also 111 hallucinations.
Fusion separates the wheat from the chaff, ranking valid requirements well ahead of hallucinations.
In the areas under the precision--recall and ROC curves, Uniform fusion alone largely surpasses every original run and configuration---by 0.142 and 0.118 over the best configuration.
Naive-Bayes weighting adds a further 0.039 and 0.052, reaching 0.864 and 0.869 while attaining useful operating points earlier.
Internal validation confirms the stability of these gains: they stay positive in at least 92\% of out-of-bag bootstrap resamples and every structured-perturbation sample.
Post-generation fusion thus turns apparent noise into a practical asset: a lightweight layer giving analysts broader coverage and a better prioritized review queue---from affordable, below-frontier models alone.
\end{abstract}


\titlepgskip=-21pt

\maketitle

\section{Introduction}
\label{sec:introduction}

\begingroup
\setlength{\parskip}{0ex plus 0.2ex minus 0.25ex}

\IEEEPARstart{L}{arge} language models (LLMs) are beginning to be explored as assistants for drafting and refining software-requirements specifications---a task that remains both linguistically demanding and labor-intensive.
Their promise is particularly compelling in cybersecurity where, abiding by security-by-design principles, engineers must translate high-level control definitions from standards such as ISO/IEC~27002 into concrete, system-specific requirements.
In this setting, LLMs can relieve analysts working under time and budget constraints of much of the drafting effort---provided that the output they must then review remains manageable.

A central challenge, however, is the diversity of LLM outputs, which stems from both stochasticity and inter-model variability.
Sampling noise causes run-to-run differences even under a fixed model and prompting scheme, and models, prompting styles, and decoding settings add further variability across configurations.
In compliance-relevant settings, this diversity matters because requirements engineering (RE) is often recall sensitive: failing to surface a relevant requirement can leave protection gaps that may not be revisited later, whereas nonrelevant (but plausible) requirements are typically easier to flag during review.
Consequently, practical adoption favors methods that improve coverage while keeping the growth in invalid requirements contained.

Most empirical studies of LLM-assisted requirements engineering still evaluate single runs only.
While this practice is reasonable for early benchmarking, it sits uneasily with the variability described above and, in particular, overlooks two salient properties of LLM behavior: stochastic diversity across runs and complementary differences across model families and prompt designs.
Recent surveys and position papers across software engineering and RE repeatedly flag stochasticity and prompt sensitivity as threats to reproducibility~\cite{Fan2023LLMSESurvey,Cheng2026GenAIREReview,Sallou2024LLMThreatsSE}, yet evaluation protocols rarely account for run-to-run variability explicitly.

Likewise, in our previous work~\cite{Yelmo2026AICyberReqGen}, which evaluated several multi-step LLM pipelines instantiating ISO/IEC~27002 controls for an LLM-evaluation-focused testbed system, we observed substantial variability across both configurations and repeated runs, and many hallucinations proved not systematic but tied to specific models or runs.
Taken together, these observations motivate a shift in perspective: instead of treating diversity purely as an evaluation nuisance, we treat it as a source of additional candidate requirements that, when combined appropriately, can improve recall while keeping the resulting increase in invalid requirements manageable.
This shift also has an economic dimension: if aggregation compensates for the limitations of any individual run, strong coverage need not depend on a single frontier model, making an ensemble of more modest---cheaper or locally operable---models an attractive alternative; and because aggregation also ranks candidates by their support, it can reach the same coverage with less review effort.

We therefore examine ensembling strictly as a post-generation strategy.
Rather than proposing new prompting pipelines or regenerating requirements, we reuse a frozen human-labeled corpus and treat each LLM run as a black-box retrieval system that returns an unranked set of candidate requirements for each cybersecurity control.
Following classical information retrieval (IR) theory, we model ensembling as a data-fusion problem over requirement sets.
Our methods include voting-based aggregation, which rewards cross-run agreement, and quality-aware weighting derived from fixed-parameter Naive-Bayes over run votes, estimating model-family reliability from repeated stochastic outputs.

We study these ensembling strategies over 12 configurations spanning four model families (Llama~3.1 405B, Qwen-2 72B, Mixtral 8$\times$22B, and GPT-4 Turbo), including multiple stochastic runs of the base generation pipelines.
Ensembles are evaluated against a gold standard (see Section~\ref{sec:materials-and-methods}) derived from expert annotation of the ISO/IEC~27002 instantiated requirements.

This research makes the following \emph{key contributions}:
\begin{enumerate}
  \item Framing post-generation ensembling as a practical response, integrable into established security-RE processes, to the stochasticity and inter-model variability of LLM outputs: treating multi-run and multi-configuration outputs as complementary signals rather than evaluation noise, and showing that fusing them acts on the \emph{order} of review and not merely its volume---even plain support-based fusion prioritizes better supported candidates, raising recall without a commensurate rise in hallucinations.
  \item Introducing a weighting scheme that treats runs as noisy observers of requirement validity, scaling each run's votes by its model family's reliability---a fixed-parameter Naive-Bayes. By discounting support from less reliable sources, it refines this ordering and reaches useful operating points at shallower review depths.
  \item Evaluating the resulting rankings with information retrieval data-fusion metrics that expose the trade-off between recall and hallucination burden across review depths, so that the fusion strategies can be compared at any compliance-relevant operating point rather than at a single fixed cut.
\end{enumerate}

The remainder of the article reviews related work (Section~\ref{sec:related-work}); describes the materials and methods underpinning our ensembling strategies (Section~\ref{sec:materials-and-methods}), the strategies themselves (Section~\ref{sec:implementation}), and the evaluation protocol (Section~\ref{sec:evaluation}); and presents the results (Section~\ref{sec:results}), their discussion (Section~\ref{sec:discussion}), and the conclusions (Section~\ref{sec:conclusion}).

\endgroup

\section{Related Work}
\label{sec:related-work}

Our work sits at the intersection of three research threads, reviewed in turn below: 1) the emerging use of large language models for requirements engineering (RE); 2) LLM-based support for defensive cybersecurity, the landscape in which our target artifact class---cybersecurity requirements---is situated; and 3) ensemble and data-fusion techniques for combining model runs or configurations, from which our approach draws its machinery.

Our previous research~\cite{Yelmo2026AICyberReqGen} highlighted two concerns that also recur in recent surveys on LLMs in software engineering and in RE: substantial output variance---across repeated runs and across models---and nontrivial hallucination rates~\cite{Fan2023LLMSESurvey,Hemmat2025LLMREReview,Cheng2026GenAIREReview}.
The present article turns that variance into the input of an ensembling step: each LLM run is treated as a black-box retrieval system, where voting-based data fusion is applied across runs and configurations.

\subsection{LLMs for Requirements Engineering}

Recent systematic reviews agree that LLM use in RE is promising but still immature: most studies remain research-prototype evaluations with scarce industrial deployment~\cite{Hemmat2025LLMREReview,Zadenoori2025LLMREReview}. Hemmat et al.\ survey natural language processing (NLP)- and LLM-based approaches across the RE lifecycle and find practical, real-world integration of LLMs in RE still underexplored~\cite{Hemmat2025LLMREReview}.
A complementary review by Zadenoori et al.\ catalogs 74 LLM-for-RE primary studies published in 2023--2024, which rely mostly on GPT-style models, are usually evaluated in controlled environments, and are seldom integrated into complex workflows~\cite{Zadenoori2025LLMREReview}.

Beyond this immaturity, the literature voices concerns about the technology itself. Cheng et al.\ focus on generative artificial intelligence (AI) for RE, reviewing 238 articles published between 2019 and 2025 and highlighting reproducibility, controllability, and governance (including safety- and security-relevant concerns) as recurrent unresolved problems~\cite{Cheng2026GenAIREReview}.
Sallou et al.\ examine the threats that LLM use poses to software-engineering research itself---including output variability and the limited reproducibility of closed, evolving models~\cite{Sallou2024LLMThreatsSE}.

These concerns materialize as practical obstacles. Norheim et al.\ review LLM applications to engineering-requirements tasks and report that, although LLMs can assist with structuring, reformulating, and checking requirements, practical use is constrained by recurring challenges: limited requirements-specific datasets, inconsistent data annotation, inadequately defined RE use cases, and adoption hurdles in engineering practice~\cite{Norheim2024REChallengesLLM}.
Practitioner-oriented discussions add that fluent output can create a misleading veneer of correctness, potentially leading to downstream errors and costly rework that offsets automation benefits~\cite{Borg2024RELLMPanel}. In security work, such residual flaws feed a familiar cost dynamic: security risk left in the software accumulates as \emph{security debt} that must later be identified, managed, and repaid~\cite{Rindell2019SecurityDebt}.

A conclusion that recurs across these studies is that human oversight remains necessary. Krishna et al.\ evaluate LLM assistance for drafting and revising software requirements specifications, finding that LLM drafts can be a productivity aid, though still short of expert-level output~\cite{Krishna2024LLMSRS}.
In the same vein, a role-prompted ChatGPT interview eliciting the requirements of a digital twin recovered about 83\% of the domain expert's reference elements while also producing items the expert marked as misinformation~\cite{Blasek2023DigitalTwinRE}.
Hymel and Johnson compare LLM-generated requirements against those of human experts and find the LLM output rated at least as aligned and complete as the experts'; even so, they conclude that humans remain essential for domain knowledge, contextual understanding, and nuanced stakeholder needs, with their role shifting toward orchestration and oversight~\cite{Hymel2025LLMvsExpertsRE}.

A further strand goes beyond oversight and casts LLMs and human engineers as collaborators, specifically around formal requirements: LLMs suggesting candidate formalizations whose soundness is then checked by formal verification~\cite{Ferrari2025FormalRELLMRoadmap}, or translating informal natural-language requirements into verifiable formal specifications and proofs~\cite{Cao2025InformalToFormal}.

More broadly, across these RE-focused studies, nearly all investigations still treat LLM output from a single run or configuration as the evaluation unit. The surveys and position papers reviewed above explicitly flag stochasticity and sampling sensitivity as threats to reproducibility~\cite{Cheng2026GenAIREReview,Sallou2024LLMThreatsSE}, yet there is comparatively little work on turning this variance into an asset---for instance, by combining multiple runs or models to increase coverage. Our previous study documented precisely this untapped complementarity in a security-requirements task: repeated runs of the same configuration overlapped only partially in the requirements they retrieved, different models agreed less still, and even the best single configuration recovered only part of what all runs found together~\cite{Yelmo2026AICyberReqGen}.
The research reported here takes up these observations as a whole: it engages several models through a generation pipeline of several steps, aligned with realistic practice; it answers the correctness and reliability concerns by ensembling multiple outputs---if LLMs are not reliable enough one at a time, fusion makes them so; and it streamlines, rather than replaces, human-in-the-loop intervention, ordering the candidate requirements so that oversight starts from the best supported ones and accounting explicitly for the cost of reviewing the hallucinations that survive.
We develop our data-fusion response to this gap from Section~\ref{sec:materials-and-methods} onward.

\subsection{LLMs for Defensive Cybersecurity}
\label{subsec:llms_for_cyber}

In parallel, there is fast-growing interest in applying LLMs to cybersecurity, with a maturity that varies markedly across the lifecycle of the artifacts addressed. Xu et al.\ provide a systematic literature review of LLMs for cybersecurity and find the surveyed work dominated by post-development, code-centric artifacts---source code, vulnerable code, bug-fix pairs---while requirements-stage activities are absent from their task taxonomy altogether~\cite{Xu2025LLM4Security}. Other surveys of the field report the same concentration---application catalogs running from vulnerability detection to secure code generation, again with no requirements-stage category~\cite{Zhang2025LLMCyberSLR,Yao2024LLMSecPrivSurvey}.
While on those post-development artifacts LLM assistance is well past the prototype stage---Divakaran and Peddinti discuss the opportunities and risks of deploying LLMs for cyber defense, for instance in interactive ``security copilot'' settings~\cite{Divakaran2025LLMCyberDefense}---for pre-development, natural-language artifacts such as cybersecurity requirements, applications remain largely exploratory.
The copilot vision itself extends to the early phases---LLMs prompted to identify and fill gaps in system specifications---though there as prospect rather than practice~\cite{Belzner2024LLMAssistedSE}.
Recent surveys synthesize the landscape of LLMs across intrusion detection, malware analysis, phishing, and incident response, identifying opportunities but also serious concerns about hallucinations, attack-surface expansion, and how easily LLMs can be misused~\cite{Hasanov2024LLMCyberReview,Jaffal2025LLMCyberSurvey}.
Motlagh et al.\ similarly catalog uses of LLMs in cyber operations, mapping defensive applications to the National Institute of Standards and Technology (NIST) cybersecurity framework: studies concentrate on the Protect and Detect functions, leaving the Identify function among the flagged research gaps---and none addresses generating the cybersecurity requirements of a project~\cite{Motlagh2025LLMCyberState}.
Taken together, these maps place our target artifact class in a region they leave blank, at the exploratory end of the field's maturity gradient; the present article contributes there the kind of prioritization layer that an early-phase security copilot would need to make its output reviewable.

More specialized reviews focus on concrete technical tasks, and two of their traits matter here. First, their objects are post-development and largely not textual: Ferrag et al.\ survey generative AI and LLMs for cybersecurity and enumerate applications in threat intelligence, secure code generation, and red-teaming~\cite{Ferrag2025GenAIForCyberSurvey}, and Sheng et al.\ review LLM-based approaches to vulnerability detection in software~\cite{Sheng2026LLMVulnDetectSurvey}---none of these targets requirements or comparable pre-development texts.
Second, they surface challenges that carry over to our study: Sheng et al.\ note strong performance on benchmark datasets but an overreliance on synthetic, isolated code and limited evaluation on complex, real-world artifacts; the security-knowledge benchmarks cataloged in recent systematic literature reviews are closed-ended, multiple-choice questions~\cite{Zhang2025LLMCyberSLR}, and benchmarks built from real-world threat-intelligence reports find LLMs handle such closed items well but fall short on open-ended analyses demanding specialized security knowledge~\cite{Ji2024SEvenLLM}.

Within this broader cybersecurity landscape, only a subset of works explicitly targets \emph{security- or privacy-related requirements} as early-lifecycle artifacts. Subahi leverages LLMs to evaluate the fulfillment of security requirements within requirements engineering~\cite{Subahi2025LLMSecReqsEval}.
Melo et al.\ review LLM applications in security requirements engineering, concluding that the research line is still emerging and lacks standardized evaluation metrics and benchmarks, which hinders reproducibility and the consolidation of results~\cite{Melo2025LLMSecurityREReview}.
Adjacent work ensembles several fine-tuned transformer models to classify requirements into functional and non-functional categories, without a security-specific focus~\cite{Alsanoosy2025LLMReqClassification}.

Our earlier work falls into this security-requirements niche. There we showed that a single, well-engineered configuration could retrieve a substantial fraction of a manually judged gold standard of ISO/IEC~27002-derived cybersecurity requirements for a given system under development, but also that 1) different runs discovered different subsets of those requirements, and 2) hallucinations were run- or model-scoped rather than systematic. Indeed, one of the prompting approaches back in that study explored a configuration that reissued the same request across four parallel runs and coalesced their outputs---aggregation without voting---raising recall but carrying every run's hallucinations into the merged result, at a cost in precision~\cite[Sec.~IV-E]{Yelmo2026AICyberReqGen}. Building on that evidence, the present article supplies what the efforts reviewed above still lack---a systematic treatment of variability across runs---by investigating whether combining multiple runs and models through principled voting-based fusion can increase recall on cybersecurity requirements without proportionally inflating hallucinations.

\subsection{Ensembling in LLMs and NLP}

Ensemble methods have a long history in machine learning and information retrieval: combining multiple imperfect predictors often yields better accuracy than relying on any single one. Classic data-fusion work in IR distinguishes between rank-based and score-based fusion and identifies the \textit{chorus effect} that makes fusion profitable~\cite{VogtCottrell1999Fusion,Wu2012DataFusionIR}, as documents appearing in many systems' results are more likely to be relevant.
Fusion schemes exploiting this effect consistently outperform individual retrieval systems in many settings: score-based rules such as CombSUM and CombMNZ~\cite{Shaw1995CombinationSearches,Wu2012DataFusionIR}, combination of Boolean---hence binary---query results~\cite{Belkin1995MultipleQueryRepresentations}, and supervised probabilistic variants like ProbFuse, which learns segment-wise probabilities of relevance from training queries and sums them across systems~\cite{Lillis2006ProbFuse}.
Beyond IR, ensemble learning is well established in software engineering more broadly, most prominently in software defect prediction~\cite{Matloob2021EnsembleDefectPrediction}; closer to our setting, data-fusion ideas themselves have recently been transferred to tasks such as code search, fusing the outputs of different search techniques as independent retrieval systems~\cite{Wang2024FusingCodeSearchers}, and bug localization, combining information-retrieval and spectrum-based signals~\cite{Le2015BugLocFusion}.

For LLMs, surveys now discuss both \textit{LLM ensembles}~\cite{Chen2025LLMEnsembleSurvey} and \textit{agentic compositions}~\cite{Jin2024LLMAgentsSurvey}.
Following the taxonomy there adopted---ensembling before, during, or after inference---we distinguish three broad families, which differ in when fusion is applied relative to generation:
\begin{enumerate}
    \item Routing approaches decide \emph{before} generation which model should handle a particular query, e.g., on expertise criteria~\cite{Lu2024Zooter}; related cost-driven cascades instead try models sequentially until an answer is accepted~\cite{Chen2024FrugalGPT}. Either way, only one model's output is ultimately used per query, so these approaches do not directly exploit the diversity across runs that we seek.
    \item Collaborative-decoding ensembles fuse models \emph{within} generation, typically at the level of token probabilities~\cite{Huang2024DeePEn}, or by selecting which model of a pool generates each reasoning step, guided by a process reward model~\cite{Park2025LEMCTS}. These methods require tight control over decoding---typically down to token-level logits or hidden states---which is often unavailable through application programming interfaces (APIs) and considerably more complex to instrument than fusing persisted outputs after the fact; they are therefore outside the scope of our study.
    \item Output-aggregation ensembles combine \emph{complete responses} after generation. Wang et al.\ propose \emph{self-consistency}, which samples multiple chain-of-thought explanations from a single model and returns the answer most consistent across samples, substantially improving accuracy on arithmetic- and commonsense-reasoning benchmarks~\cite{Wang2023SelfConsistency}.
    Jiang et al.'s LLM-Blender first collects outputs from several base LLMs, uses a learned pairwise ranker to score them, and then prompts a generator model to synthesize a final answer that fuses high-scoring candidates~\cite{Jiang2023LLMBlender}.
    Li et al.\ and other recent work show that agreement-based voting over independently sampled answers can serve as a strong, training-free ensemble baseline~\cite{Li2024MoreAgents}.
    Output-level stacking of several LLMs has also been applied to security-relevant classification tasks such as phishing detection~\cite{Nasser2025StackingLLMPhishing}.
\end{enumerate}

Both IR data fusion and LLM output ensembling address a similar problem through similar methods---combining the outputs of several imperfect systems into a single, better result. From output-aggregation \emph{ensembling}---the LLM-side term of art for the overall paradigm---we inherit the working conditions: complete responses combined after generation, with no access to model internals. From IR data fusion we take the machinery---the concrete combination rules operating inside that paradigm; indeed, each classical scheme reviewed above anticipates one element of our design (Section~\ref{sec:implementation}): CombSUM, additive score fusion itself; the combination of binary Boolean evidence, the vote counting underlying both of our strategies; and ProbFuse, the supervised, probability-derived computation of the reliability weights specific to our Naive-Bayes strategy. What requires adaptation is the structure of the inputs: whereas IR data fusion merges the \emph{ranked} lists returned by several search systems, each of our LLM runs behaves as an unranked retrieval system returning a set of candidate requirements for a given control and system specification, so a run's contribution reduces to a binary vote per candidate. Under this reading the chorus effect transfers directly---agreement across runs signals validity. The transfer has qualitative support in the source study's consistency analysis: introducing different models increased diversity (reduced pairwise Jaccard similarity), and both inter-model and same-model consistency were higher on valid requirements than on hallucinations~\cite{Yelmo2026AICyberReqGen}. We study two voting-based fusion strategies: Uniform fusion, which counts votes and thereby operationalizes the chorus effect, and Naive-Bayes fusion, which further weights each vote by the estimated reliability of the model that produced it. Unlike output-aggregation methods built on auxiliary trained models, our strategies train no additional model---the Naive-Bayes weights are fitted directly from adjudicated data---and operate purely post hoc on the runs' persisted textual outputs.

\section{Materials and Methods}
\label{sec:materials-and-methods}

This study builds on the results of a previous evaluation of LLM-augmented cybersecurity requirements generation~\cite{Yelmo2026AICyberReqGen}, in which several LLMs repeatedly drafted system-specific cybersecurity requirements from standard-derived control templates, and experts subsequently adjudicated every candidate they produced.
In this section, we restate that task setting and the frozen, expert-labeled corpus reused here, define the contingency-table metrics used over its universe of candidate requirements, and cast each LLM execution of the task as a black-box retrieval system whose outputs are fused into a single ranked review queue using the strategies of Section~\ref{sec:implementation}.

\subsection{Task Description and Source Corpus}
\label{sec:task-description}

\begin{figure}[t]
  \centering
  \includegraphics[width=\linewidth]{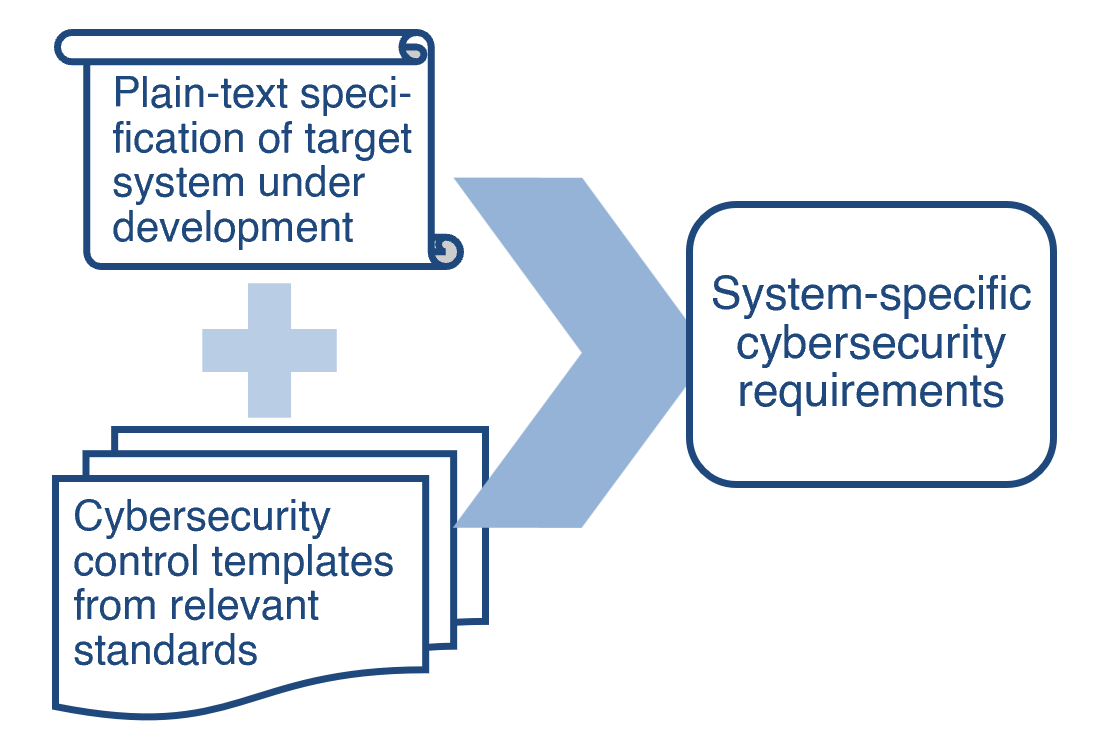}
  \caption{Standards-driven cybersecurity requirements generation task (adapted from~\cite{Yelmo2026AICyberReqGen}).}
  \label{fig:task_functional_view}
\end{figure}

As summarized in Fig.~\ref{fig:task_functional_view}, the source study addressed an LLM entrusted with the task of drafting system-specific cybersecurity requirements, provided with 1) a natural-language description of the target system under development and 2) a parameterized cybersecurity control template derived from ISO/IEC~27002:2022, to be instantiated into concrete requirements for the target system. The problem is therefore not free-form ideation, but the contextualization of generic control intent into system-specific requirements.

The execution of the task combines several steps in an LLM-based pipeline built with LangChain that concatenates applicability checks, the mapping of template placeholders to domain elements, the generation of free-text requirement instances, and JavaScript Object Notation (JSON) formatting.
Ten cybersecurity control templates derived from ISO/IEC~27002:2022 and annotated with placeholders were used as input; their parameters denote elements such as user roles, sensitive data stores, or protected activities that must be bound to concrete elements in the specification of the system under development.
The target system was AI4I4 (Automated Identification and Data Capture for Industry~4.0), a realistic project addressing AIDC technologies in the logistics processes of an automotive factory, created as a fit-for-purpose research testbed to benchmark LLM performance in this area.
Its documentation is an approximately 5300-word natural-language specification covering the relevant domain concepts, roles, data stores, and use cases needed to ground cybersecurity requirements.

The elicitation process goes through each control template in the input and generates the requirements that follow from its instantiation over the target system's elements. Each execution of this process, which we refer to as a \emph{run} of the task, produces its own set of candidate requirements. Runs are executed under a \emph{configuration} that fixes model, prompt pipeline, and LLM settings; even repeated runs of the same configuration differ, through stochastic decoding.
As for the models, their selection was inherited from the source study. Four were used, one from each of four families: three open-weights and operable on local infrastructure (Llama~3.1 405B, Qwen-2 72B, Mixtral 8$\times$22B), and one proprietary, accessed remotely as a commercial service (GPT-4 Turbo).
This selection comprises models affordable as of today and provides a fitting testbed for the article's central question: whether ensembles of such affordable (and hence not frontier-class) models can produce better results than any of their individual members.
For each model, we reuse from the source study one base configuration, executed as four independent stochastic runs, and two additional single-run configurations with different prompting/pipeline and LLM-setting choices, as summarized in Table~\ref{tab:configs}.
The detailed prompt, pipeline, and hyperparameter settings were reported in the source study~\cite{Yelmo2026AICyberReqGen} and in its own reproducible research package~\cite{Yelmo2025RRPackage}.

\begin{table}[t]
  \caption{Source-study configurations and their runs reused in the ensemble study}
  \label{tab:configs}
  \setlength{\tabcolsep}{3pt}
  \begin{threeparttable}
  \begin{tabular}{p{75pt} p{70pt} p{80pt}}
    \hline
    Model                & Run ID(s)\,\tnote{a}                & Variant description \\
    \hline
    Llama 3.1 405B       & L0$_{\mathrm{i}}$--L0$_{\mathrm{iv}}$ & Base chain          \\
    Llama 3.1 405B       & L1                                  & Cooler chain        \\
    Llama 3.1 405B       & L2                                  & Warmer chain        \\
    Qwen-2 72B           & Q0$_{\mathrm{i}}$--Q0$_{\mathrm{iv}}$ & Base chain          \\
    Qwen-2 72B           & Q1                                  & Cooler chain        \\
    Qwen-2 72B           & Q2                                  & Warmer chain        \\
    Mixtral 8$\times$22B & M0$_{\mathrm{i}}$--M0$_{\mathrm{iv}}$ & Base chain          \\
    Mixtral 8$\times$22B & M1                                  & Cooler chain        \\
    Mixtral 8$\times$22B & M2                                  & Warmer chain        \\
    GPT-4 Turbo          & G0$_{\mathrm{i}}$--G0$_{\mathrm{iv}}$ & Base chain          \\
    GPT-4 Turbo          & G1                                  & Cooler chain        \\
    GPT-4 Turbo          & G2                                  & Warmer chain        \\
    \hline
  \end{tabular}
  \begin{tablenotes}
    \item[a] Subscripts $\mathrm{i}$--$\mathrm{iv}$ index the four stochastic runs of each base configuration; the variant configurations contributed one run each.
  \end{tablenotes}
  \end{threeparttable}
\end{table}

For each run, the source corpus records the candidate requirements generated by the LLM.
The control templates provide an ex-ante grouping of the candidates: each candidate is associated with the control template from which it was elicited, although our analysis below generally aggregates over controls.
Raw generated requirements were then:
1) pooled across runs, collecting all candidate statements produced by the source runs;
2) decomposed, when a statement bundled several atomic requirements; and
3) canonicalized, so that semantically equivalent statements were represented once, mapping paraphrases or repeated rediscoveries to a single item.
The pool also includes requirements elicited by human analysts following the same control-template instantiation process.
During corpus construction, each canonical candidate requirement was expert-adjudicated as valid or hallucinated, with valid candidates mapped to existing human-elicited requirements when they expressed the same normative intent, or added as new nonredundant valid requirements when appropriate.
Conversely, some valid requirements may have no support from any LLM run.

This yields a frozen judged corpus over which all post hoc ensembling and evaluation are performed, without any further interaction with the underlying LLMs.
The full requirement text behind these candidates is distributed with the source study's reproducible research package~\cite{Yelmo2025RRPackage}, against which the candidate identifiers used throughout this article resolve; our own deposit records the reused run composition and the ensembling analysis (see the Appendix).

\subsection{Judged Universe and Contingency-Table Metrics}
\label{sec:supporting-dataset}

Let \(\mathcal{U}\) denote the universe of candidate requirements.
Following the human adjudication process described above, it is partitioned into \(\mathcal{V}\), the gold-standard valid requirements, and \(\mathcal{H}\), the hallucinated candidates.
In the version of the corpus used in this article, \(|\mathcal{U}|=183\) and \(|\mathcal{V}|=72\).
The universe is delimited by this study's participating sources, the same membership rule applying to valid and hallucinated candidates alike: \(\mathcal{U}\) comprises the canonical candidates emitted by the 24 reused runs together with the requirements elicited by the human analysts.
Consequently, \(\mathcal{V}\) is the source study's gold standard re-pooled over this universe's participating runs: of its 76 published valid requirements, the 6 contributed only by exploratory source-study configurations are excluded here (alternative prompting pipelines, and a coalesced multi-branch variant that would have skewed the run set toward one model), plus we include 2 further requirements surfaced only by the repeated runs---which the source study analyzed for consistency but did not pool into its gold standard---for 72 in total; the hallucinated candidates of the excluded configurations likewise do not enter \(\mathcal{U}\).

Let \(\mathrm{Runs}\) be the set of LLM runs.
For each run \(r\in\mathrm{Runs}\), let \(\mathcal{U}_r\subseteq\mathcal{U}\) denote the set of canonical candidate requirements generated by that run.
Each run thus defines a binary vote
\begin{equation}
y_r(u)=
\begin{cases}
1, & u\in \mathcal{U}_r,\\
0, & u\notin \mathcal{U}_r,
\end{cases}
\qquad u\in\mathcal{U}.
\end{equation}

For any selected set \(A\subseteq\mathcal{U}\), such as the output \(\mathcal{U}_r\) of a run, the counts of the contingency table (true/false positives/negatives) are
\begin{equation}
\begin{gathered}
\mathrm{TP}(A)=|A\cap\mathcal{V}|,
\qquad
\mathrm{FP}(A)=|A\cap\mathcal{H}|,\\
\mathrm{FN}(A)=|\mathcal{V}\setminus A|,
\qquad
\mathrm{TN}(A)=|\mathcal{H}\setminus A|.
\end{gathered}
\end{equation}
True negatives are thus defined only within the finite set \(\mathcal{H}\) of hallucinated candidates produced during corpus construction---not over every invalid statement one could conceivably write: our goal is to evaluate how well alternative fusion strategies prioritize the candidates that were actually produced, pooled, and judged.

From these counts, we compute the recall or true-positive rate (TPR), the precision or positive predictive value (PPV), the specificity or true-negative rate (TNR), and the fall-out or false-positive rate (FPR):
\begin{equation}
\begin{aligned}
\mathrm{TPR}(A)&=\mathrm{TP}(A)/(\mathrm{TP}(A)+\mathrm{FN}(A)),\\
\mathrm{PPV}(A)&=\mathrm{TP}(A)/(\mathrm{TP}(A)+\mathrm{FP}(A)),\\
\mathrm{TNR}(A)&=\mathrm{TN}(A)/(\mathrm{TN}(A)+\mathrm{FP}(A)),\\
\mathrm{FPR}(A)&=1-\mathrm{TNR}(A)=\mathrm{FP}(A)/(\mathrm{TN}(A)+\mathrm{FP}(A)).
\end{aligned}
\end{equation}

This represents a closed-world evaluation in the Cranfield/Text REtrieval Conference (TREC) tradition: systems are compared over a fixed test collection and a fixed set of relevance judgments~\cite{Cleverdon1967Cranfield,VoorheesHarman2005TREC,Sanderson2010TestCollections}.
The retrieval analogy is deliberate, with two peculiarities. First, there is a single, fixed query---in effect, ``the requirements instantiating these controls on this system''---rather than a stream of varying information needs. Second, the collection is the judged universe \(\mathcal{U}\) itself, pooled from the outputs of all runs and from the human-elicited requirements, and then manually adjudicated. Within it, the valid requirements \(\mathcal{V}\) play the role of the relevant items, and the hallucinated candidates in \(\mathcal{H}\) that of the nonrelevant ones.

None of these ratios is informative in isolation: recall can be driven arbitrarily high by selecting more candidates, at the expense of precision and specificity, and conversely. We therefore use two scalar summaries that aggregate both sides of this trade-off, weighted to encode how costly a missed valid requirement is relative to a spurious candidate in our review setting.
The first is the \(F_{\beta}\)-measure~\cite{vanRijsbergen1979IR,Christen2023FMeasureReview}, a weighted harmonic average of recall and precision,
\begin{equation}
F_{\beta}=(1+\beta^2)\,\mathrm{PPV}\,\mathrm{TPR}/\bigl(\beta^2\,\mathrm{PPV}+\mathrm{TPR}\bigr).
\end{equation}
For \(\beta=1\), this is the usual \(F_1\)-score.
In this study, we use \(F_2\), which places four times more weight on recall than on precision.

The second weighted summary is a weighted Youden index.
The usual Youden index summarizes the balance between sensitivity and specificity~\cite{Youden1950Index} as \(J=\mathrm{TPR}+\mathrm{TNR}-1\). We use instead the weighted form~\cite{Li2013WeightedYouden}
\begin{equation}
J_w=
2\bigl(
w\,\mathrm{TPR}
+
(1-w)\,\mathrm{TNR}
\bigr)-1,
\end{equation}
which reduces to \(J\) when \(w=1/2\).
We adopt the cost-sensitive formulation of optimal operating points on the receiver operating characteristic (ROC) curve~\cite{GreinerPfeifferSmith2000,PerkinsSchisterman2006ROC} and compute
\begin{equation}
w=\rho\,\pi/\bigl(\rho\,\pi+(1-\pi)\bigr),
\end{equation}
where \(\pi\) is the prevalence of relevant elements in our universe \(\pi=|\mathcal{V}| / |\mathcal{U}|\),
and \(\rho\) is a scenario-dependent false-negative to false-positive cost ratio.
With \(\rho=4\) (a 4:1 preference to avoid false negatives over false positives), this gives \(w=0.722\).
\(F_2\) and \(J_w\) aggregate different ratios, but both encode the same utility-oriented methodological preference: in this task, preserving valid requirements is more important than minimizing the number of candidates that experts must reject, as omitted controls can create security gaps, compliance risk, and expensive late rework, whereas a hallucinated candidate can be more easily discarded during review.

Single-run outputs are used as original-output baselines.
In addition, configuration-level summaries are provided as baselines that average the numerical evaluations of the runs sharing the same model, prompt/pipeline, and LLM settings.

\subsection{Ensembling as Voting-Based Data Fusion}
\label{sec:ensembling-strategies}

\begin{figure*}[!t]
  \centering
  \includegraphics[width=0.85\textwidth]{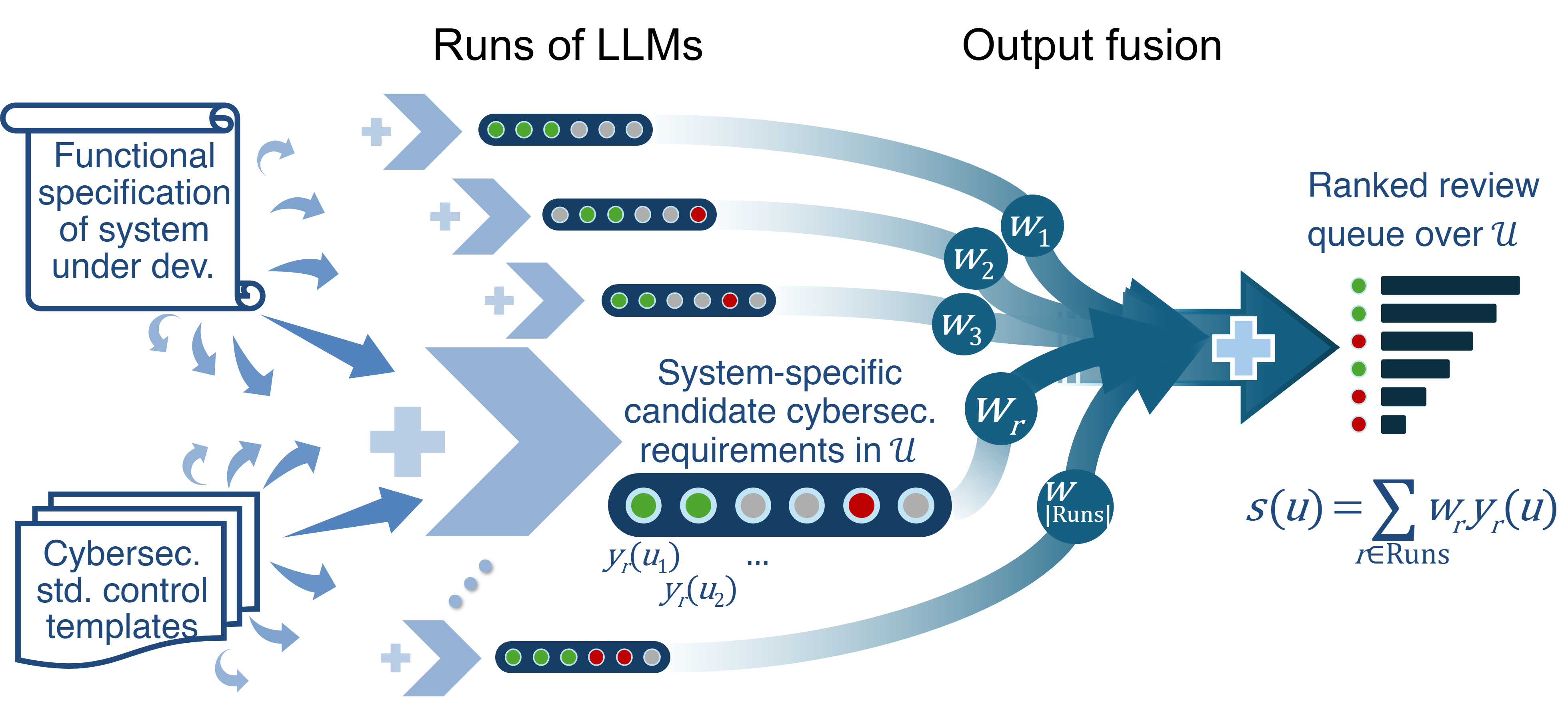}
  \caption{Post-generation ensembling: votes cast by parallel LLM runs are fused, through per-run weights, into a single ranked review queue over the judged universe \(\mathcal{U}\).}
  \label{fig:ensembling_process}
\end{figure*}

Post-generation ensembling may be treated as voting-based data fusion over the outputs of the black-box runs, as shown in Fig.~\ref{fig:ensembling_process}.
Each run casts a binary vote \(y_r(u)\) on each candidate \(u\in\mathcal{U}\), and these votes are combined into a fused score \(s(u)=\sum_{r\in\mathrm{Runs}} w_r\,y_r(u)\) through per-run weights \(w_r\).
Unlike the runs' unranked output sets, the fused output is a single ranked list over \(\mathcal{U}\), obtained by sorting candidates by decreasing score; this ranking is the article's main object of analysis.
In a practical review workflow, candidates would then be inspected from the top of the ranking until the available review budget or an acceptance threshold is reached.
Every such stopping point selects a prefix---a set to which the contingency-table metrics above apply directly---so a ranking defines a family of operating points, which Section~\ref{sec:evaluation} assesses both individually and across all review depths.

The underlying intuition is the data-fusion ``chorus effect'': support for a candidate from several runs, especially runs of independent systems, provides evidence of relevance~\cite{Lee1997EvidenceCombination,VogtCottrell1999Fusion}. At the same time, not all sources are equally reliable: giving more influence to runs from models that are empirically better at separating valid requirements from hallucinations may produce more faithful results.
Section~\ref{sec:implementation} formalizes the concrete scoring rules and the weight choices that realize these intuitions as the two fusion strategies evaluated in this article: Uniform fusion and Naive-Bayes fusion.
\section{Implementation}
\label{sec:implementation}

In this section, we introduce the fusion strategies, which convert the unranked outputs of the source LLM runs into a single scored ranking over the judged candidate universe \(\mathcal{U}\).

\subsection{Weight-Based Ensembling}
\label{sec:weight-based-ensembling}

For each candidate requirement \(u\in\mathcal{U}\), the weight-based ensemble uses only the binary votes \(y_r(u)\) cast by the source runs \(r\in\mathrm{Runs}\)---the indicators of membership in each run's output set \(\mathcal{U}_r\), as defined in Section~\ref{sec:supporting-dataset}.
A fusion strategy assigns a weight \(w_r\) to each run and computes the aggregated score of the candidate requirement
\begin{equation}
  s(u)
  =
  \sum_{r\in\mathrm{Runs}} w_r\,y_r(u).
  \label{eq:ensemble-score-run}
\end{equation} 
Candidates are ranked by decreasing \(s(u)\)---the classical CombSUM rule from information-retrieval data fusion~\cite{FoxShaw1994Combination,Lee1997EvidenceCombination}, specialized to binary run-level evidence~\cite{Belkin1995MultipleQueryRepresentations}.
Because the inputs are unranked sets, \eqref{eq:ensemble-score-run} fuses agreement rather than ranks: the score is built post hoc from the generators' votes, with no per-run ordering to merge.

Although \eqref{eq:ensemble-score-run} allows arbitrary run weights, the strategies reported here do not estimate idiosyncratic weights for individual stochastic runs.
The reason is that runs produced by the same configuration are repeated executions of the same model, prompt pipeline, and LLM settings; thus, assigning them different learned weights would mainly capture accidental variation in a particular corpus realization.
Hence, all the strategies give equal weight to runs from the same configuration.
In particular, the two strategies evaluated in the article are even coarser: Uniform fusion gives the same weight to all runs, whereas Naive-Bayes fusion gives the same weight to all runs produced by the same model.

\subsection{Fusion Strategies}
\label{sec:implementation-strategies}

The first strategy we present is Uniform fusion, which assigns the same weight to every run,
\begin{equation}
  w_r^{\mathrm{unif}} = 1/|\mathrm{Runs}|,
  \qquad
  \forall r\in\mathrm{Runs}.
  \label{eq:uniform-weight}
\end{equation}
The normalization constant has no effect on the ranking, but makes the score scale independent of the number of runs.
Uniform fusion therefore ranks candidates by the amount of support they receive across runs: a candidate's score is simply the fraction of runs that generated it.

The second strategy is Naive-Bayes (NB) fusion, which is based on estimating how much a vote from each model should shift the evidence toward candidate validity.
Let \(Y_u\in\{0,1\}\) denote the latent validity label of candidate \(u\), with \(Y_u=1\) for \(u\in\mathcal{V}\) and \(Y_u=0\) for \(u\in\mathcal{H}\).
For each model \(m\), let \(\mathrm{Runs}_m\subseteq\mathrm{Runs}\) be the subset of runs produced by that model.
The Naive-Bayes model treats the votes of runs in \(\mathrm{Runs}_m\) as conditionally independent repeated measurements given \(Y_u\), with shared true-positive and false-positive rates:
\begin{equation}
\begin{aligned}
  \mathrm{TPR}_m
  &=
  \Pr\!\left(y_r(u)=1 \mid Y_u=1,\, r\in\mathrm{Runs}_m\right),\\
  \mathrm{FPR}_m
  &=
  \Pr\!\left(y_r(u)=1 \mid Y_u=0,\, r\in\mathrm{Runs}_m\right).
\end{aligned}
\label{eq:nb-rate-model}
\end{equation}

Both assumptions---conditional independence and shared rates across the runs of a model---are practical approximations: runs of the same model are not independent, as prompt and pipeline reuse induces dependence among their votes.
Nonetheless, such assumptions do not hamper the validity of our results, as they are used only to derive this strategy's weight vector; the empirical evaluation in Sections~\ref{sec:evaluation} and~\ref{sec:results} assesses the ranking induced by that vector, whether or not the assumptions hold.
The result is a practical weighting rule, not necessarily an optimal one: better weightings exist for a given corpus, and our evaluation includes one, fitted numerically, as a benchmark for the attainable headroom (Section~\ref{sec:evaluation}).

Because the corpus provides expert labels for all \(u\in\mathcal{U}\), the rates are estimated directly, as
\begin{equation}
\begin{aligned}
  \widehat{\mathrm{TPR}}_m
  &=
  \frac{
    \sum_{r\in\mathrm{Runs}_m}
    \sum_{u\in\mathcal{V}} y_r(u)
    + 1/2
  }{
    |\mathrm{Runs}_m|\,|\mathcal{V}| + 1
  },\\
  \widehat{\mathrm{FPR}}_m
  &=
  \frac{
    \sum_{r\in\mathrm{Runs}_m}
    \sum_{u\in\mathcal{H}} y_r(u)
    + 1/2
  }{
    |\mathrm{Runs}_m|\,|\mathcal{H}| + 1
  }.
\end{aligned}
\label{eq:nb-rate-hat}
\end{equation}
The \(+1/2\) terms in the numerators and \(+1\) in the denominators are the result of Jeffreys regularization (equivalent to a \(\mathrm{Beta}(1/2,1/2)\) prior for each Bernoulli rate) to avoid zero or unit rate estimates, which would otherwise produce infinite log-odds weights.
The same half-count correction is standard in probabilistic information retrieval weighting, notably in Robertson--Sparck Jones relevance weighting~\cite{RobertsonSparckJones1976RelevanceWeighting}.

The corresponding model weight is the log-likelihood-ratio contribution of an observed vote toward requirement validity:
\begin{equation}
  w_m^{\mathrm{NB}}
  =
  \log
  \left(
    \frac{
      \widehat{\mathrm{TPR}}_m
      \left(1-\widehat{\mathrm{FPR}}_m\right)
    }{
      \widehat{\mathrm{FPR}}_m
      \left(1-\widehat{\mathrm{TPR}}_m\right)
    }
  \right).
  \label{eq:nb-logodds}
\end{equation}
Each run \(r\in\mathrm{Runs}_m\) is then assigned \(w_r=w_m^{\mathrm{NB}}\) in \eqref{eq:ensemble-score-run}.
The weight is signed and deliberately not clipped at zero: a positive value indicates that a vote from model \(m\) increases the posterior log-odds of validity, whereas a negative value would mark that model's votes as evidence against validity.
In this corpus, all estimated weights are positive.
This produces a supervised weighted-voting rule: a candidate supported by models with high estimated recall over valid requirements and low estimated fall-out over hallucinations receives more score mass.
Under the model of \eqref{eq:nb-rate-model}, the CombSUM score admits an exact posterior reading.
By Bayes' rule, the posterior log-odds of validity given the full vote profile is the prior log-odds plus one log-likelihood-ratio term per run: \(\log(\mathrm{TPR}_m/\mathrm{FPR}_m)\) for a run of model \(m\) that voted for \(u\), and \(\log\bigl((1-\mathrm{TPR}_m)/(1-\mathrm{FPR}_m)\bigr)\) for one that did not---in the closed universe \(\mathcal{U}\), a non-generated candidate is an informative zero vote, not a missing judgment, so silence also counts as evidence.
Rewriting each term as the silence contribution plus \(y_r(u)\,w_m^{\mathrm{NB}}\)---the weight of \eqref{eq:nb-logodds} is precisely the gap between the vote and silence contributions---and noting that every candidate thereby collects the silence contributions of all runs, the posterior log-odds equals \(s(u)\) plus a candidate-independent constant: the prior log-odds plus \(\sum_m |\mathrm{Runs}_m|\log\bigl((1-\mathrm{TPR}_m)/(1-\mathrm{FPR}_m)\bigr)\).
Ranking by \(s(u)\) is therefore ranking by the Naive-Bayes posterior \(\Pr\!\left(Y_u=1 \mid \{y_r(u)\}_{r\in\mathrm{Runs}}\right)\): absent votes need no separate weight, and the prevalence prior, shifting all candidates equally, leaves the ranking unchanged.
The reading is specific to the judged universe on which the rates are estimated: prospective use on a different system under development would apply these weights as fixed, best-effort estimates---the fixed-weight regime whose robustness the perturbation analysis of Section~\ref{sec:evaluation-stability} probes---and the dataset dependence this entails is discussed in Section~\ref{sec:limitations}.

This formulation is closely related to the well-known Dawid--Skene model for noisy observers~\cite{DawidSkene1979}, where each observer has a confusion matrix and the typically unknown true labels are inferred jointly with the observer reliabilities through expectation--maximization. Here, each LLM run acts as a noisy observer that either retrieves or fails to retrieve a candidate, and the expert labels are known, so the observer reliabilities are estimated directly from the judged corpus as shown in \eqref{eq:nb-rate-hat}.
\section{Evaluation Protocol}
\label{sec:evaluation}

The evaluation compares the fusion rules with each other and with the original LLM outputs over the judged candidate universe \(\mathcal{U}\), attending to two complementary concerns: the coverage of valid requirements, captured by recall, and the hallucination burden, captured by precision and fall-out (equivalently, specificity), giving more relevance to the former through the combined yet recall-laden criteria \(F_2\) and \(J_w\). Each of these metrics varies along the review depth \(k\) (the @\(k\) operating points), while the global metrics---average precision (AP) and the area under the ROC curve (ROC AUC)---summarize whole rankings by respectively integrating the combined precision--recall and ROC curves.
Because vote-based scores produce frequent ties, all metrics are computed as tie-aware expected values. In addition, single runs and configuration averages are themselves evaluated with equivalent metrics, which place them on the same axes as the fused rankings.
The results are framed with respect to reference curves and subject to internal-validation analyses to interpret whether the observed differences are robust to plausible changes in the evidence available for evaluation.

\subsection{Ranked Outputs and @\headk{} Operating Points}
\label{sec:evaluation-metrics}

Both the ensembles and the single runs are interpreted as retrieval outputs over \(\mathcal{U}\), but with different score structures.
A single run \(r\) produces an unranked set \(\mathcal{U}_r\): its generated candidates, those with vote \(y_r(u)=1\), form one tied block above its non-generated remainder, with \(y_r(u)=0\). An ensemble, by contrast, induces a meaningful ranked list by sorting candidate requirements by the fusion score \(s(u)\) from \eqref{eq:ensemble-score-run}.

For a ranked list, inspecting the first \(k\) candidates defines a rank cutoff, commonly denoted by the suffix @\(k\) in information retrieval.
Changing \(k\) changes the operating point of the retrieval: larger \(k\) usually increases coverage, measured by recall, but also admits more hallucinated candidates, lowering precision and specificity.
For a cutoff \(k\), the \(\mathrm{Top}_k\) prefix set induces a contingency table (with \(\mathrm{TP}@k\), \(\mathrm{FP}@k\), \(\mathrm{FN}@k\), and \(\mathrm{TN}@k\)).
Therefore the operating-point metrics (such as \(P@k\), \(R@k\), \(\mathrm{Specificity}@k\), \(F_2@k\), and \(J_w@k\)) --- generically, \(\mathcal{M}@k\) for an evaluation metric \(\mathcal{M}\) --- can be computed by applying the formulas from Section~\ref{sec:supporting-dataset} to the set \(A = \mathrm{Top}_k\).
Degenerate cases are handled with common-sense defaults matching their natural limits: a prefix containing no true positive has \(F_2@k=0\) (resolving the \(0/0\) form), and a universe lacking valid items (possible in the resampled universes of the stability analysis, Section~\ref{sec:evaluation-stability}) yields recall, AP, and ROC AUC equal to \(0\), while one lacking hallucinated candidates yields specificity and ROC AUC vacuously equal to \(1\).

In our analysis, we will emphasize two fixed review depths.
The first is \(k_M=\max_{r\in\mathrm{Runs}} |\mathcal{U}_r|\), i.e., a review budget the size of the largest original LLM output (in this corpus, \(k_M=40\)).
The second is \(k_R = |\mathcal{V}|\) (named after the usual \(R\)-precision convention where \(R\) denotes the number of relevant items), which represents the earliest rank at which full recall could in principle be achieved (in this corpus, \(k_R=72\)).

We also report the maximum values of \(F_2@k\) and \(J_w@k\), together with the first rank at which each maximum is attained.
These maxima summarize the best operating point under the chosen recall-sensitive utility view, and the associated rank indicates how early it is reached.

\subsection{Curves and Global Ranking Metrics}
\label{sec:evaluation-curves}

As explained above, each operating-point metric \(\mathcal{M}@k\) yields a curve rather than a single value: plotted as an explicit function of the review depth, it shows how the quality of the review set evolves as the operating point deepens.
Sweeping \(k\) also parametrizes --- implicitly, with \(k\) running along each curve --- two complementary views that integrate the trade-offs between the individual metrics.

The precision--recall (PR) curve plots \(P@k\) against \(R@k\).
This curve makes visible how many valid requirements are recovered (recall or TPR) for a given density of useful review items (precision or PPV).
The ROC curve plots \(\mathrm{TPR}@k\) (recall) against \(\mathrm{FPR}@k\) (fall-out).
The PR and ROC views are complementary: precision ties the PR view to the prevalence of valid candidates, whereas the ROC view, built from per-class rates, is insensitive to prevalence~\cite{Fawcett2006ROC,DavisGoadrich2006PRROC}.
Each combined criterion is a function of the two coordinates of one of these planes: \(F_2\) of precision and recall, \(J_w\) of TPR and FPR. Every point of the corresponding curve therefore fixes the criterion's value, so \(F_2\) can be read directly off the PR curve and \(J_w\) off the ROC curve.

The overall quality of each of these curves can, in turn, be captured by a single metric of its own:
\begin{itemize}
     \item AP can be understood as the area under the conventional stepwise precision--recall curve induced by ranking relevant items (slightly different from the geometric area under a visually plotted line segment curve~\cite{NISTTREC2007Measures}). It summarizes whether valid requirements are retrieved early and at high precision.
     \item ROC AUC is the area under the empirical ROC curve. It summarizes pairwise separability between valid and hallucinated candidates, as it equals the probability that a randomly chosen valid requirement receives a higher score than a randomly chosen hallucinated candidate.
\end{itemize}

\subsection{Tie Handling}
\label{sec:evaluation-ties}

Discrete vote-based scores create frequent ties.
Hence, a rank cutoff may fall inside an equal-score block; likewise, global metrics such as AP or ROC AUC may depend on how tied candidates are internally ordered.
Four standard remedies exist: report only score-block-level values, break ties with an arbitrary secondary key, simulate random tie-breaks, or compute analytical expectations over random permutations within each tie block.
Following tie-aware IR evaluation, we use the last~\cite{McSherryNajork2008TiedScores}, whose closed-form expectations cover the operating-point metrics used here (precision, recall, F-scores) plus the global AP.

After sorting by score, let the ranking be partitioned into maximal tie blocks. We take a uniformly random permutation over each tied block, which contains \(t\) items of which \(v\) are valid. When a prefix cuts through such a block, each additional position inside the block contributes an expected valid mass of \(v/t\) true positives and an expected hallucinated mass of \((t-v)/t\) false positives.
Thus, \(\mathrm{TP}@k\) and \(\mathrm{FP}@k\) are estimated by these fractional expected counts, and since every operating-point metric at a fixed \(k\) is an affine function of the counts, replacing them in its formula yields its exact expected value.
This avoids arbitrary document-ID effects, is deterministic and reproducible, and reports the expected performance implied by the score information actually available to the fusion strategy.
Note that the reported maximum-attaining rank for a given metric \(\mathcal{M}\) (e.g., \(F_2\) or \(J_w\)) is therefore the first \(k\) maximizing the expected curve, i.e., \(\arg\max_k \mathbb{E}[\mathcal{M}@k]\), rather than the expectation of a random maximizing rank, \(\mathbb{E}[\arg\max_k \mathcal{M}@k]\).
Ties affect the estimate only when the cut falls inside a tie block: blocks lying entirely within the top-\(k\) prefix contribute their exact counts, so @\(k\) values at block boundaries involve no expectation at all.

The global metrics AP and ROC AUC, in contrast, are not affine in the counts and require dedicated treatment under the same tie model.
Expected AP is computed from closed-form block contributions under uniform random ordering within tied blocks: in a block occupying ranks \(b,\dots,b+t-1\) with \(v\) valid items among its \(t\) members and preceded by \(R_0\) valid items, the position \(b+j\) (for \(j=0,\dots,t-1\)) contributes
\begin{equation}
\frac{v}{t}\cdot
\frac{R_0+1+j\,\frac{v-1}{t-1}}{b+j}
\end{equation}
in expectation to the sum of precisions at valid ranks, whose total over all blocks, divided by \(|\mathcal{V}|\), is the expected AP.
Expected ROC AUC is computed from the pairwise preference probability, with each valid--hallucinated pair tied in the same block receiving half credit~\cite{Fawcett2006ROC,Bamber1975ROCArea}.

\subsection{Comparability Metrics for Single Runs and Configuration Averages}
\label{sec:evaluation-comparability}

The curves and global metrics defined above describe ranked ensemble outputs. To place single runs on the same axes as the fusion rankings, each run is evaluated as a binary retrieval system over \(\mathcal{U}\): its induced two-level score \(y_r(u)\) ranks the generated candidates as one tied block above the non-generated remainder, and the run's equivalent metrics are those of the resulting two-block ranking.

This does not imply that the LLM produced an inherent ordering: the two-level score just encodes membership in \(\mathcal{U}_r\).
For this binary scorer, ROC AUC reduces to \(\mathrm{AUC}_r =
 {(\mathrm{TPR}_r+\mathrm{TNR}_r)}/{2}\), and \(\mathrm{AP}_r\) is the expected AP of the two-block ranking under the tie model of Section~\ref{sec:evaluation-ties}.

In addition, configuration summaries are used as baselines for repeated runs of the same configuration; the equivalent metrics defined next place these averages on the same axes as well. These should be read as descriptive baselines that smooth outlier runs, never as the output of some additional ensemble.
Let \(\mathrm{Runs}_c\) be the runs belonging to configuration \(c\).
For each configuration \(c\), the ROC coordinates are micro-averaged over the common assessment pool:
\begin{equation}
\mathrm{TPR}_c =
\frac{
\sum_{r\in\mathrm{Runs}_c}\mathrm{TP}(\mathcal{U}_r)
}{
|\mathrm{Runs}_c|\,|\mathcal{V}|
},
\;\;
\mathrm{FPR}_c =
\frac{
\sum_{r\in\mathrm{Runs}_c}\mathrm{FP}(\mathcal{U}_r)
}{
|\mathrm{Runs}_c|\,|\mathcal{H}|
}.
\end{equation}
To display the same configuration point in PR space, precision is reconstructed backward from population prevalence \(\pi=|\mathcal{V}|/|\mathcal{U}|\):
\begin{equation}
\mathrm{PPV}_c =
\frac{\pi\,\mathrm{TPR}_c}
     {\pi\,\mathrm{TPR}_c + (1-\pi)\,\mathrm{FPR}_c}.
\end{equation}
The remaining operating-point metrics, including \(F_2\) and \(J_w\), are computed from these micro-averaged rates; the equivalent operating point for these configuration summaries is placed at the mean output size
\begin{equation}
\bar{k}_c =
\frac{1}{|\mathrm{Runs}_c|}
\sum_{r\in\mathrm{Runs}_c} |\mathcal{U}_r|.
\end{equation}
Configuration ROC AUC is likewise obtained from configuration TPR and TNR through the same binary-scorer identity as for single runs, \((\mathrm{TPR}_c+\mathrm{TNR}_c)/2\), not by integrating a ROC curve.
These averaging choices are deliberate: since each run has its own output size and hence its own precision denominator, the plain average of per-run precisions and recalls is not the precision--recall point of any actual retrieval, whereas micro-averaging the counts places the configuration at the operating point of its pooled contingency table.

AP, in contrast, has no pooled counterpart --- there is no single configuration-level ranking whose AP it would describe --- so we report the macro-average of single-run AP values:
\begin{equation}
\mathrm{AP}_c =
\frac{1}{|\mathrm{Runs}_c|}
\sum_{r\in\mathrm{Runs}_c} \mathrm{AP}_r .
\end{equation}
The configuration's PR point and its \(\mathrm{AP}_c\) are thus deliberately separate aggregates --- micro-averaged rates versus a macro-average over runs: absent a pooled configuration ranking, neither is derived from the other.

\subsection{Benchmarks: Reference Curves and Normalized Uplift}
\label{sec:evaluation-reference-curves}

To contextualize the numerical differences observed between one strategy and another, and between these and single runs, we frame the computed curves within four reference curves, each bounding the results in a different sense:

\begin{enumerate}
\item The \emph{feasible region}, imposed by a dataset with \(|\mathcal{V}|\) valid requirements and \(|\mathcal{H}|\) hallucinated candidates. This region represents the geometry of the finite judged universe, as no ranking can retrieve more than \(k\) valid items by rank \(k\) or more than \(|\mathcal{V}|\) valid requirements in total; hence full precision is unattainable at review depths beyond \(|\mathcal{V}|\), bounding the values each operating-point metric \(\mathcal{M}@k\) can take at any given \(k\).

\item The \emph{expected random ranking}, computed as a single tie block containing all candidates in \(\mathcal{U}\).
Given population prevalence \(\pi=|\mathcal{V}|/|\mathcal{U}|\), the expected counts at rank \(k\) are \(\mathbb{E}[\mathrm{TP}@k]=k\pi\) and \(\mathbb{E}[\mathrm{FP}@k]=k(1-\pi)\). Consequently, \(\mathbb{E}[P@k]=\pi\), \(\mathbb{E}[R@k]={k}/{|\mathcal{U}|}\), and \(\mathbb{E}[\mathrm{FPR}@k]={k}/{|\mathcal{U}|}\).
The random PR curve is therefore horizontal at the prevalence level \(\pi\), whereas the random ROC curve follows the chance diagonal.
Derived quantities such as \(F_2@k\) and \(J_w@k\) are computed from these expected point metrics.

\item The \emph{pattern oracle}.
Candidate requirements can be grouped into \emph{pattern blocks}: two candidates share a block when every configuration gives them the same vote count, i.e., when their configuration-level vote patterns are identical.
Candidates within the same pattern block will receive the same score under any fusion rule that only considers vote counts from each configuration.
That is, if a pattern block contains both valid and hallucinated candidates, no weighting scheme that uses only those voting features can separate them.
The pattern oracle produces the ranking that sorts these unbreakable pattern blocks by their observed internal precision (fraction of valid candidates), from highest to lowest, while leaving candidates inside each block tied.
It therefore estimates the headroom available if vote patterns were exploited perfectly, without introducing any additional information.

\item The \emph{numerical linear-fusion benchmark}, built by searching over nonnegative configuration-weight vectors in the CombSUM family, refining trial weight vectors by coordinate ascent on the simplex.
At each step, one coordinate is swept over a grid, the remaining weights are rescaled, the induced ranking is rescored, and the move is accepted only if it improves the target tie-aware objective.
This reference is heuristic rather than a certified optimum, because the objective is nonsmooth and depends on score-induced tie blocks.
Nevertheless, it estimates the headroom potentially attainable by a linear combination of vote counts just by changing fusion weights.
Comparing a reported fusion rule to this numerical benchmark isolates suboptimal weighting, whereas comparing that benchmark to the pattern oracle isolates irreducible vote-pattern ties.
\end{enumerate}

All four accompany the fusion strategy curves in the compound figure of Section~\ref{sec:results} (Fig.~\ref{fig:f2-pr-roc-compound}).

For each global metric \(\mathcal{M}\) (AP or ROC AUC), we report the normalized uplift of any given ranking approach, as
\begin{equation}
\mathrm{uplift}_{\mathcal{M}}[\mathrm{approach}]
=
100\,
\frac{
\mathcal{M}[\mathrm{approach}]-\mathcal{M}[\mathrm{random}]
}{
\mathcal{M}[\mathrm{benchmark}]-\mathcal{M}[\mathrm{random}]
},
\end{equation}
where the approach may be a fusion strategy or a baseline such as the best single configuration. This expresses how much of the observed gap between random ordering and the numerical linear-fusion benchmark is recovered by a simple, interpretable fusion rule.

\subsection{Internal Validation and Stability}
\label{sec:evaluation-stability}

The full-data results are apparent performance estimates on the same judged corpus used to estimate weights. Therefore, we need to probe internal validity and assess whether the relative conclusions about fusion rules are stable to plausible changes in what the evaluation treats as evidence.
The primary internal-validation analysis is a requirement-level out-of-bag (OOB) bootstrap with refitting: we draw 200 bootstrap replicates, each a sample of 183 candidates drawn with replacement from the 183 candidate requirements in \(\mathcal{U}\), and collapse duplicate in-bag candidates to preserve a set-valued retrieval universe.
Data-dependent fusion strategy weights are then refitted on the unique in-bag candidates and evaluated on the out-of-bag complement.
The same out-of-bag complement is used for all strategies in a replicate, thus differences are paired.
The weight \(w\) of \(J_w\), in contrast, keeps its full-data value without refitting, both here and in the perturbation samples below, so the metric stays on a single utility scale and paired differences remain comparable.

This analysis estimates whether the relative advantage of a strategy persists when the learned weights are applied to unseen candidates from the same judged universe; transfer to unseen control templates is not separately validated.
The resulting contrasts are reported in the frame of estimation statistics: each is summarized by the magnitude of the paired difference in the metric's own units --- themselves probability-scale quantities --- and by how that magnitude varies across replicates, where percentile bands are interpreted as descriptive internal-validation ranges, not as classical confidence intervals~\cite{EfronTibshirani1993Bootstrap,PolitisRomanoWolf1999Subsampling}.
Because the replicates overlap and do not form an external validation sample, test statistics computed naively over them are anti-conservative~\cite{NadeauBengio2003GeneralizationError}.

The OOB procedure is related in spirit to \(m\)-out-of-\(n\) bootstrap subsampling ideas, but is not identical: we draw \(n\) times with replacement and then evaluate on the unsampled complement, so the effective in-bag size is random.
In any case, the results obtained from this OOB procedure are intentionally conservative.
Deduplication leaves an expected unique in-bag fraction of \(1-(1-1/183)^{183}\approx 63\%\) of the corpus; empirically, a mean of about \(116\) unique in-bag and \(67\) out-of-bag candidates over the \(200\) replicates. Performance estimated after training on such reduced bootstrap samples is known to be pessimistically biased --- the learning-curve effect behind the ``.632'' correction of the leave-one-out bootstrap~\cite{EfronTibshirani1993Bootstrap} --- so OOB values are best read as lower bounds on the corresponding full-data performance.

The complementary stability analysis is a fixed-weight perturbation analysis: we perturb the evidence while keeping the already-fitted fusion weights fixed.
We use combined leave-one-out perturbations over runs and the ex-ante control slices---the per-template groups of candidate requirements noted in Section~\ref{sec:task-description}---dropping one control slice and one run at a time, in a jackknife-like fashion.
That way, we check whether results are driven by particular control templates or by idiosyncratic runs.

Where available, we also report analytic intervals that support the interpretation of selected scalar comparisons: DeLong-style inference for ROC AUC~\cite{DeLong1988ROC} and Wald approximations for fixed-threshold Youden-type operating-point quantities~\cite{Li2013WeightedYouden}.
These intervals are full-data, per-comparison instruments, and they rest on an independence idealization: the judged candidates are treated as independent draws, whereas candidates cluster by the control template that elicited them --- a clustering probed empirically by the control-slice perturbations rather than modeled analytically.
The internal-validation argument itself is carried by the paired OOB and perturbation analyses, on a different ground: within each replicate or perturbation sample, all strategies are evaluated on the identical resampled universe, so the paired differences preserve the dependence between strategies --- a pairing argument, not a correction for candidate clustering.
\section{Result Analysis}
\label{sec:results}
This section presents the empirical evidence for the contributions of Section~\ref{sec:introduction}: instantiating the protocol of Section~\ref{sec:evaluation}, we compare the rankings of the two fusion strategies (Uniform and Naive-Bayes) against each other and the original outputs---single runs and configuration averages as unranked baselines---first across review depths (Section~\ref{sec:results-effects}) and then by global ranking quality (Section~\ref{sec:results-ranking-quality}).
Under both views, the findings unfold in a two-stage pattern: the mere agreement among runs already improves the order in which candidate requirements would be reviewed---Uniform fusion yields substantial gains over every original output---and the reliability-aware Naive-Bayes weighting refines that ordering further, reaching useful operating points earlier.

\subsection{Ensembling Effects Across Review Depths}
\label{sec:results-effects}

\begin{figure}[p]
  \setlength{\abovecaptionskip}{3pt}
  \setlength{\belowcaptionskip}{0pt}
  \centering
  \includegraphics[width=\columnwidth,height=\dimexpr\textheight-2.5\baselineskip\relax,keepaspectratio]{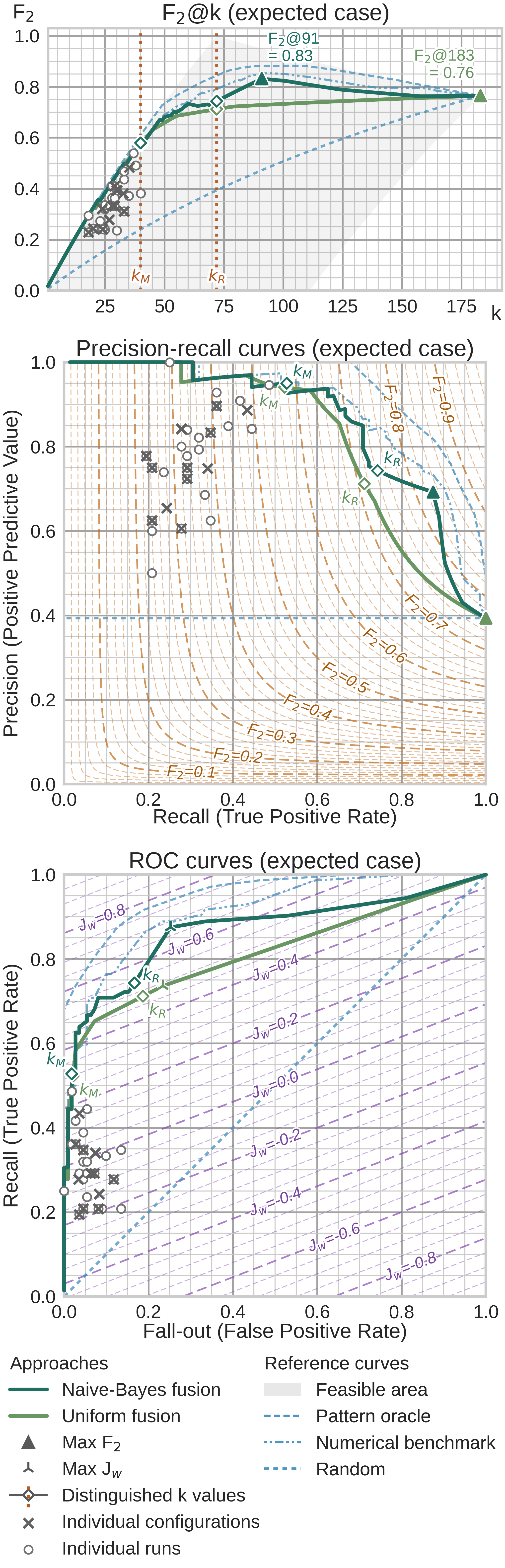}
  \caption{Evaluation of the fusion strategies (Uniform and Naive-Bayes) and the original outputs (single runs and configuration averages).}
  \label{fig:f2-pr-roc-compound}
\end{figure}

As the analyst descends through the ranked list of candidate requirements produced by a given fusion strategy, performance metrics vary, trading exhaustiveness against tolerance to hallucinations.
Fig.~\ref{fig:f2-pr-roc-compound} compares, over the judged universe, the retrieval behavior of the original LLM outputs and of the two fusion strategies.
The upper panel plots \(F_2@k\) against the review depth \(k\), providing an effort-aligned summary of recall and precision; the PR and ROC panels then show the same ranked outputs in terms of review yield and class separation.
All results use the expected tie handling defined in Section~\ref{sec:evaluation-ties}.

The two fusion strategies appear as curves swept by the operating point \(k\), with dedicated markers flagging the maximum \(F_2\) and \(J_w\) values each curve attains.
The original outputs appear as points: individual runs, and configuration averages placed at their mean output size \(\bar{k}_c\) with micro-averaged coordinates (Section~\ref{sec:evaluation-comparability}).
The distinguished review depths \(k_M\) and \(k_R\) appear as vertical references in the upper panel and are marked on the curves in the other two.
The four reference curves of Section~\ref{sec:evaluation-reference-curves} frame the comparison: the shaded feasible region, the expected random ranking, the pattern oracle, and the numerical linear-fusion benchmark.
For the latter, each panel draws the boundary given by the search's best solution for the global objective underlying its respective view: AP for the \(F_2@k\) and PR panels, and ROC AUC for the ROC panel.
Finally, the background contours in the PR and ROC panels trace constant-\(F_2\) and constant-\(J_w\) levels, so that our two recall-sensitive criteria can be read directly off both views.

One feature of the figure deserves clarification: the long final straight segment of the Uniform curve stems from its coarse support-count score.
The large group of candidates supported by exactly one run forms one equal-score block, within which valid and hallucinated mass accumulates at a constant fractional rate under tie handling, producing straight segments in the \(F_2@k\) and ROC views.
Naive-Bayes weighting, by contrast, separates part of this block according to the reliability of the supporting models, thus yielding a finer-grained ordering.
Note that the numerical benchmark is optimized for a global metric, not for pointwise dominance, so a strategy curve may locally rise above its boundary.

As shown in all three panels, Uniform fusion already yields a substantial gain before any reliability information is introduced, and the improvement runs along every dimension of interest: its fused ranking offers better precision at a given recall and better recall at a given precision than the original outputs, together with lower fall-out and higher \(F_2\) and \(J_w\) at comparable review depths.
Table~\ref{tab:cutoffs-summary} quantifies this at the two workload-aligned operating points.
Under a conservative analyst workload, we may consider that the reviewer inspects at most the largest list that a single run might emit (\(k_M=40\)); there, Uniform fusion markedly improves all of \(P@k_M\), \(R@k_M\), and \(\mathrm{FPR}@k_M\) relative to the best configuration, and consequently also both recall-laden utility metrics \(F_2@k_M\) and \(J_w@k_M\). A deeper representative operating point is the number of valid requirements in the gold standard (\(k_R = 72\)).

In turn, Naive-Bayes fusion adds a smaller but consistent improvement over Uniform:
across the practically relevant range of review depths \(k\), reliability weighting shifts the PR and ROC curves toward higher precision and lower fall-out for a given recall, higher recall for a given precision, and better combined recall-favoring metrics (higher \(F_2\) and \(J_w\)).

\begin{table}[!b]
  \centering
  \small
  \caption{Metrics of the best configuration and the fusion strategies at distinguished operating points}
  \label{tab:cutoffs-summary}
  \setlength{\tabcolsep}{2.4pt} 

  \newcommand{\vcell}[1]{\shortstack{#1}}

  \begin{tabular}{@{} l l c c@{\hskip 0.5pt} c@{\hskip 1pt} c c @{} } 
    Approach
      & \shortstack{Operating\\point \(k\)}
      & \diaghead{\(P@k\)}
      & \diaghead{\(R@k\)}
      & \diaghead{\(\mathrm{FPR}@k\)}
      & \diaghead{\(F_2@k\)}
      & \diaghead{\(J_w@k\)} \\
    \hline
    \shortstack[l]{Best config. (L0)}
      & {\rule{0pt}{2.6ex}\(\bar{k}_{\mathrm{L0}} = 35.25\)}
      & \vcell{0.887} & \vcell{0.434} & \vcell{0.036} & \vcell{0.483} & \vcell{0.163} \\

    \shortstack[l]{Naive-Bayes}
      & \vcell{\(k_M = 40\)}
      & \vcell{0.950} & \vcell{0.528} & \vcell{0.018} & \vcell{0.579} & \vcell{0.308} \\

    Uniform
      & \vcell{\(k_M = 40\)}
      & \vcell{0.941} & \vcell{0.523} & \vcell{0.021} & \vcell{0.574} & \vcell{0.299} \\
    \hline\hline

    \shortstack[l]{Naive-Bayes}
      & \vcell{\(k_R = 72\)}
      & \vcell{0.743} & \vcell{0.743} & \vcell{0.167} & \vcell{0.743} & \vcell{0.536} \\

    Uniform
      & \vcell{\(k_R = 72\)}
      & \vcell{0.712} & \vcell{0.712} & \vcell{0.187} & \vcell{0.712} & \vcell{0.480} \\
    \hline
  \end{tabular}
\end{table}

\subsection{Ensembling Gains in Global Ranking Quality}
\label{sec:results-ranking-quality}

Table~\ref{tab:full-data-summary} confirms the operating-point picture with global metrics that summarize effectiveness over the whole ranking.
\begin{table*}[!t]
  \caption{%
    Global metrics for the best configuration (L0) and the fusion strategies: AP and ROC AUC with normalized uplifts (progress from the expected random ordering toward the numerical linear-fusion benchmark), and the maximum \(F_2\) and \(J_w\) operating points%
  }
  \label{tab:full-data-summary}
  \centering
  \setlength{\tabcolsep}{3pt}

  \sbox{\fullDataSummaryBox}{%
    \small
    \begin{tabular}{@{}lcccccccc@{}}
      \hline
      &
      \multicolumn{2}{c}{Average Precision} &
      \multicolumn{2}{c}{ROC AUC} &
      \multicolumn{2}{c}{\(F_2\) operating point\,\textsuperscript{a}} &
      \multicolumn{2}{c}{\(J_w\) operating point\,\textsuperscript{a}}
      \\
      \cline{2-9}
      Approach
      & Value & Uplift
      & Value & Uplift
      & \(k\) & Value
      & \(k\) & Value
      \\
      \hline
      Best configuration (L0)
      & 0.684 & 56.8\%
      & 0.699 & 48.4\%
      & 35.25 & 0.483
      & 35.25 & 0.163
      \\
      Uniform fusion
      & 0.825 & 86.2\%
      & 0.817 & 76.9\%
      & 183 & 0.764
      & 79 & 0.489
      \\
      Naive-Bayes fusion
      & 0.864 & 94.1\%
      & 0.869 & 89.6\%
      & 91 & 0.831
      & 91 & 0.679
      \\
      \hline
    \end{tabular}%
  }

  \usebox{\fullDataSummaryBox}

  \vspace{2pt}

  \begin{minipage}{\wd\fullDataSummaryBox}
    \scriptsize\raggedright
    \textsuperscript{a} For the best configuration, $F_2$ and $J_w$ are evaluated at its mean native output size $\bar{k}_{\mathrm{L0}}=35.25$. For the fusion strategies, the table reports the maximum value over $k$ and the first rank at which it is attained.
  \end{minipage}

\end{table*}

If we normalize the results to the gap between random ordering and the numerical linear-fusion benchmark, the best original configuration covers roughly half of it (56.8\% for AP, 48.4\% for ROC AUC), whereas Uniform fusion recovers 86.2\% of the AP gap and 76.9\% of the ROC AUC gap.

The chorus effect thus emerges as a strong ranking signal: repeated support across runs prioritizes valid requirements more effectively than any original configuration alone.

Naive-Bayes fusion further improves the results by weighting support according to model reliability: normalized uplifts reach 94.1\% for AP and 89.6\% for ROC AUC.
These gains over Uniform are moderate in magnitude but consistent across the PR
and ROC views, indicating that reliability weighting improves both early
placement and pairwise separation.

The maxima of \(F_2@k\) still call for a qualified reading on two axes at once, as the difference is as much one of review depth as of peak attainable quality: Naive-Bayes not only reaches a higher maximum (0.831 against 0.764) but also does so much earlier (\(k=91\)), whereas Uniform peaks only after the candidate pool has been exhausted and the whole judged universe has effectively been admitted (\(k=183\)).
For \(J_w\) the contrast in value (0.679 against 0.489, both reached at comparable review depths) reflects a better recall--fall-out trade-off under the chosen false-negative cost preference.
The benefit of reliability-aware weighting therefore lies not merely in
higher attainable maxima, but in a more useful ordering that reaches a favorable
recall--hallucination balance within a shallower review.

\section{Discussion: Impact, Stability, and Scope}
\label{sec:discussion}

This section reads the results of Section~\ref{sec:results} from three complementary angles, answering the questions a practitioner would raise before acting on them: what the improvements buy in practice (coverage, review effort, and fit with security requirements engineering processes; Section~\ref{sec:impact-results}); whether the differences are stable under resampling and structured perturbations (Section~\ref{sec:results-internal-validity}); and within what scope the findings should be interpreted (Section~\ref{sec:limitations}).

\subsection{Impact of Results}
\label{sec:impact-results}
The results presented in Section~\ref{sec:results} have two main implications.
First, when each model run is viewed as a noisy observer that recovers a partial set of valid requirements, post-generation ensembling improves the recall achieved.
Although the best single configuration (L0) covers on average less than half of the gold-standard requirements, the ensemble of the outputs produced by the different models, configurations, and runs recovers all 72 of them.

Second, ensembling turns output variability into ranked evidence.
Run-to-run stochasticity and inter-model differences manifest as dispersion in which valid requirements each run recovers and which hallucinations it introduces; the ordering induced by post-generation ensembling with appropriate weights allows reaching the same recall level with lower review budgets (i.e., going through shorter-length candidate lists), as candidate requirements supported by several runs---or by more reliable sources---surface earlier.
Merely pooling the partial outputs already yields a more complete space of candidate requirements, and reviewing extra candidates to prune false positives may be an acceptable price for the added coverage---but this pruning entails a review cost, so the binding constraint in a human-in-the-loop workflow may eventually be a \emph{fixed} number of items an analyst can review.
Operationally, the question is not only ``does ensembling generate more items?'' but also ``does it prioritize better items within the prefix that is actually reviewed?''
As shown by ROC AUC, ensembling induces meaningful rankings where valid requirements are more consistently placed ahead of hallucinations; likewise, the density of hallucinations encountered early for the same ranking prefix length is reduced.
Consequently, review effort is more productive: the budget concentrates on higher confidence candidates, and the choice of \(k\) lets the reviewer trade coverage against hallucination burden.
Taken together, the two implications weigh most in a recall-sensitive activity such as cybersecurity requirements engineering, where a missed requirement costs more than the manual discarding of a spurious addition: pooled coverage supplies the missing valid requirements, and the fused ordering makes recovering them affordable within a bounded review budget.

Within this picture, Uniform fusion is a finding in its own right: plain support aggregation, with no reliability information at all, already outperforms every original configuration and captures most of the attainable gain over random ordering (Section~\ref{sec:results-ranking-quality}).
Repeated support across runs is thus informative by itself---a rediscovered candidate is a better review target than one produced only once.
Naive-Bayes fusion adds a further, reliability-weighted gain on top of this chorus effect by discounting support from less reliable models.
Comparisons confined to peak utility values understate that refinement, as they miss the ordering advantage: Naive-Bayes not only reaches a higher maximum \(F_2\) than Uniform, but it also does so at roughly half the review depth (Section~\ref{sec:results-ranking-quality}).
Max \(F_2\) thus becomes the weakest of the internally validated contrasts (Section~\ref{sec:results-internal-validity}): shared peak values conceal the ordering advantage of the earlier ascent.

The granularity of each fusion rule also delimits what it can order.
Uniform fusion cannot distinguish among the candidates supported by a single run, which form the large equal-score block noted in Section~\ref{sec:results-effects}; the per-model Naive-Bayes weights partially resolve this block by ordering single-vote candidates according to the reliability of their supporting model.
The indistinguishability that remains is only partly a limitation of the weighting: candidates with identical vote patterns are inseparable to any vote-based rule, and the pattern oracle of Section~\ref{sec:evaluation-reference-curves} quantifies exactly that bound.

The applicability conditions of the approach are modest: several LLM runs producing overlapping candidate sets that can be canonicalized into comparable units and judged.
Post-generation ensembling fuses and prioritizes what the runs produce; it does not recover valid requirements that appear in none of them.

These properties translate into support for established security requirements engineering processes.
A concrete example is Security Quality Requirements Engineering (SQUARE), a methodology from the Software Engineering Institute that organizes the process into nine steps, from agreeing on definitions to a final requirements inspection~\cite{Mead2005SQUARE}.
The source study already envisioned implementing SQUARE as a cooperative process between LLMs and human analysts~\cite{Yelmo2026AICyberReqGen}.
Within that vision, the results reported here bring concrete improvements to three of the methodology's steps:
\begin{itemize}
    \item At elicitation (Step~6), fusing several runs enlarges the pool of standards-linked candidates beyond what any single run recovers.
    \item At prioritization (Step~8), the reliability-weighted scores add an evidence-of-validity criterion---the strength of cross-run support---to the benefit and effort the step weighs.
    \item At inspection (Step~9), the fused ranking hands reviewers a prioritized queue whose depth can be matched to the available inspection budget, spending it on the best supported candidates first---while final acceptance remains a human decision.
\end{itemize}

Finally, the models ensembled here are not frontier models as of 2026---which points to an opportunity: ensembles of cheaper, lower tier, or open-weights, locally operable models may offer a practical route to results otherwise sought from a single frontier model, a direction already visible in systems that assemble open-source models into ensembles surpassing strong proprietary ones~\cite{Wang2025MixtureOfAgents} and in multi-LLM systems shipped as a single product~\cite{SakanaFugu2026}.
And even though the source study framed its model rankings as temporal snapshots rather than enduring hierarchies~\cite{Yelmo2026AICyberReqGen}, we posit that the relative advantage of post-generation ensembling over its own constituent runs is the durable part of these results: agreement among imperfect generators remains informative whoever the generators are.
The frontier may advance; the result holds.

\subsection{Internal Validation and Construct Stability}
\label{sec:results-internal-validity}

The full-corpus improvements of Section~\ref{sec:results} are apparent estimates, potentially sensitive to the particular corpus of requirements; we therefore examine whether the observed differences persist under out-of-bag resampling and structured perturbations of the available evidence.

Following the estimation-statistics frame of Section~\ref{sec:evaluation-stability}, the mean paired difference is the effect estimate of each contrast, the p05--p95 range describes the spread of the difference across replicates, and the win rate records how often it is positive; analytic confidence intervals complement these for the two metrics whose distribution theory is available.

\begin{figure}[t]
  \centering
  \includegraphics[width=\columnwidth,keepaspectratio]
    {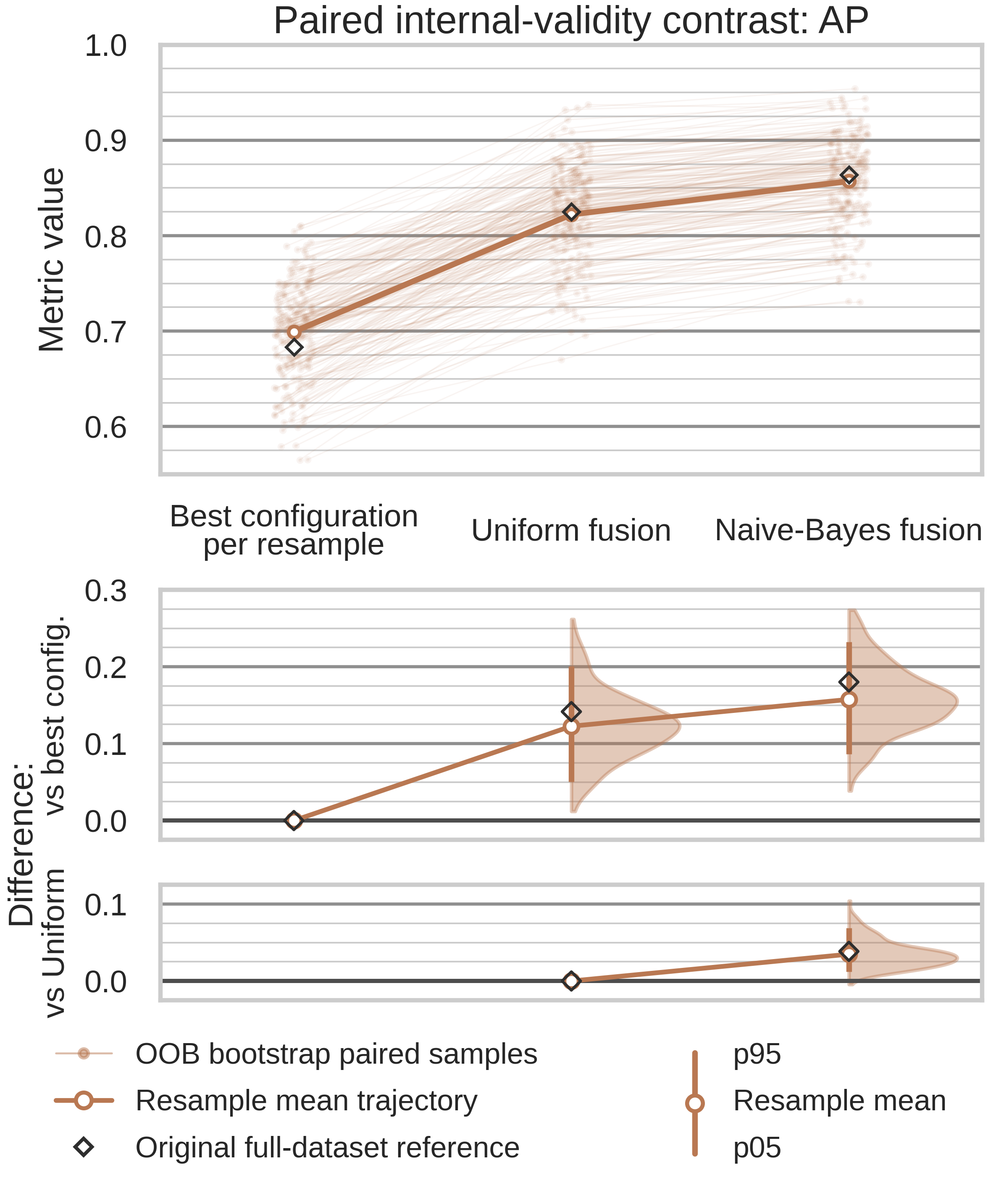}
  \caption{%
    Paired Cumming estimation plot for AP under the requirement-level OOB
    bootstrap.
  }
  \label{fig:paired-internal-validity-ap}
\end{figure}

Fig.~\ref{fig:paired-internal-validity-ap} provides internal-validation evidence through a paired Cumming estimation plot~\cite{Cumming2012NewStatistics,Ho2019EstimationGraphics} that presents paired contrasts of AP under the out-of-bag bootstrap procedure of Section~\ref{sec:evaluation-stability}.
The upper lane shows a paired trajectory plot (slopegraph): each faint line connects the AP values of a given out-of-bag test sample (making the paired structure explicit) across the three approaches; namely, the best single configuration (reselected within each replicate), Uniform fusion, and Naive-Bayes fusion.
The superimposed trajectory connects the mean value for each approach.
The two lower lanes show the paired difference distributions (deltas) relative to 1) the best
configuration and 2) Uniform fusion, each summarized by a half-violin with its p05--p95 bar and mean.

Table~\ref{tab:oob-paired-deltas} extends the estimation-plot summary to the
four principal metrics, presenting the improvement in two steps: Uniform
fusion against the best single configuration, and Naive-Bayes fusion against
Uniform fusion.
For each comparison it reports the full-data difference; the OOB mean
difference with its descriptive p05--p95 range and its win rate over the 200
overlapping replicates; and, where distribution theory permits, an analytic
95\% confidence interval (CI) and two-sided \(p\)-value.
The win rate is the proportion of replicates in which the difference is
positive---the empirical probability of superiority \(\Pr(d>0)\)~\cite{McGrawWong1992CommonLanguage},
with exact ties not counted as wins.
The Monte Carlo error of each reported OOB mean difference---\(\mathrm{SD}/\sqrt{B}\), where SD is the standard deviation of the paired deltas and \(B=200\) is the number of replicates---stays below 0.007: the reported means are stable with respect to the finite number of replicates, a statement about simulation precision rather than population inference.

\begin{table*}[t]
  \caption{%
    Paired internal-validation differences among the three approaches%
  }
  \label{tab:oob-paired-deltas}
  \centering
  \setlength{\tabcolsep}{4pt}

  \sbox{\oobPairedDeltasBox}{%
    \small
    \begin{tabular}{llrrrrrrr}
    \hline
    & &
    & \multicolumn{3}{c}{OOB bootstrap}
    & \multicolumn{3}{c}{Analytic}
    \\
    \cmidrule(lr){4-6}\cmidrule(lr){7-9}
    Comparison & Metric
    & Full-data \(\Delta\)
    & Mean \(\Delta\)
    & p05--p95
    & Win rate\,\textsuperscript{a}
    & \(\Delta\)
    & 95\% CI
    & \(p\)\,\textsuperscript{a}
    \\
    \hline
    \multirow{4}{*}{\shortstack[l]{Uniform fusion vs.\\best configuration}}
    & AP
    & 0.142 & 0.123 & [0.051, 0.201] & 100.0\% & --- & --- & ---
    \\
    & ROC AUC
    & 0.118 & 0.111 & [0.048, 0.174] & 100.0\%
    & 0.083\,\textsuperscript{b} & [0.020, 0.145] & 0.010
    \\
    & Max \(F_2\)
    & 0.281 & 0.285 & [0.172, 0.385] & 100.0\% & --- & --- & ---
    \\
    & Max \(J_w\)\,\textsuperscript{c}
    & 0.326 & 0.346 & [0.205, 0.496] & 100.0\%
    & 0.241\,\textsuperscript{b} & [0.071, 0.410] & 0.005
    \\
    \hline
    \multirow{4}{*}{\shortstack[l]{Naive-Bayes vs.\\Uniform fusion\,\textsuperscript{d}}}
    & AP
    & 0.039 & 0.035 & [0.012, 0.069] & 99.5\% & --- & --- & ---
    \\
    & ROC AUC
    & 0.052 & 0.050 & [0.012, 0.096] & 99.0\%
    & 0.052 & [0.010, 0.094] & 0.015
    \\
    & Max \(F_2\)
    & 0.067 & 0.060 & [0.000, 0.114] & 92.5\% & --- & --- & ---
    \\
    & Max \(J_w\)\,\textsuperscript{c}
    & 0.190 & 0.160 & [0.054, 0.294] & 100.0\%
    & 0.190 & [0.055, 0.326] & 0.006
    \\
    \hline
    \end{tabular}%
  }

  \usebox{\oobPairedDeltasBox}

  \vspace{2pt}

  \begin{minipage}{\wd\oobPairedDeltasBox}
    \scriptsize\raggedright
    \textsuperscript{a}A row's OOB win rate and its analytic \(p\) quantify
    different comparisons and must not be read as one.\\
    \textsuperscript{b}Compared against the best single run for the metric
    (paired DeLong for ROC AUC, paired Wald for max
    \(J_w\)~\cite{WenzelZapf2013,ZhouQin2013PairedYouden}), rather than
    the best configuration used by the OOB columns.\\
    \textsuperscript{c}Each side is evaluated at its own maximizing rank
    \(k\), so a paired difference generally compares different operating
    points.\\
    \textsuperscript{d}All columns share the Uniform baseline; the analytic
    \(\Delta\) thus equals the full-data \(\Delta\).
  \end{minipage}

\end{table*}

The first comparison isolates the chorus effect: plain Uniform fusion already improves on the best single configuration on all four metrics in every replicate.
Across them, the mean advantage amounts to roughly 3--4.5 times the replicate-to-replicate standard deviation of the paired deltas.
Since the baseline is reselected on each replicate's test sample, a choice that favors it, the comparison is conservative; the advantage is therefore not an artifact of fixing one configuration chosen once on the full dataset.

The second comparison is more demanding, because Uniform already exploits the chorus effect; it thus isolates the added value of reliability-aware weighting.
The Naive-Bayes--Uniform differences in AP, ROC AUC, and max \(J_w\) remain positive throughout the central 90\% OOB range, with win rates of 99.5\%, 99.0\%, and 100\%, respectively.
Max \(F_2\) is the weakest of the four contrasts, at a 92.5\% win rate, its p05 of zero reflecting a residue of replicates in which the two strategies reach the same maximum; the difference in maximum value and in the review depth at which it is attained is discussed in Section~\ref{sec:results-ranking-quality}.

The total advantage of Naive-Bayes fusion over the best single configuration---the first difference lane of Fig.~\ref{fig:paired-internal-validity-ap}---amounts, for AP, to a mean OOB difference of 0.158, positive in every replicate.

Analytic full-data intervals are available for ROC AUC and the Youden operating-point metrics, and provide a complementary view (Table~\ref{tab:oob-paired-deltas}).
All four lie entirely above zero: for both metrics, Uniform fusion improves on the best single run---a stricter baseline, since a configuration's strongest run can only exceed that configuration's average---and Naive-Bayes fusion improves on Uniform fusion.
The intervals for the Youden metrics are pointwise traditional (Wald) confidence intervals~\cite{Li2013WeightedYouden} evaluated at the selected operating-point values for \(k\). They treat sensitivity and specificity as independent binomial proportions estimated on the relevant and nonrelevant candidates, respectively, and they do not adjust for the preceding maximization over \(k\), which also leaves the dependence induced by selecting the operating point on the same data unmodeled~\cite{BantisNakasReiser2014}.

Finally, the joint control-slice/run perturbation (Section~\ref{sec:evaluation-stability}) complements OOB refitting by applying the full-corpus weight vector after removing one control slice and one run at a time.
Under these fixed-weight perturbations, Naive-Bayes fusion outperforms Uniform fusion on all four metrics in every one of the 240 samples.
The conclusions about global ordering quality and weighted recall--specificity
balance are therefore not driven by one particular control slice or one
idiosyncratic run.

\subsection{Scope Limitations and Validity Constraints}
\label{sec:limitations}

This exploratory study addresses only the post-generation aggregation stage: we reuse a frozen, expert-labeled corpus of generated outputs and evaluate IR-inspired voting and quality-weighted fusion rules over canonicalized requirement sets, holding fixed everything upstream---the elicitation, prompting, and generation that produced those outputs.
Its conclusions therefore pertain to the behavior of \emph{output-aggregation} ensembles, not to the intrinsic capabilities of any model, prompt library, or multi-step generation pipeline.
It consequently inherits the external-validity constraints of the source study---a single standard and system, one language, and one expert team's adjudication. However, the difference in construct attenuates them: our claims concern the \emph{relative} effect of ensembling with respect to base performance, not the absolute performance of any model, so factors that shift the whole field of runs up or down weigh far less on the within-study contrasts we report than they would on an absolute claim.
Moreover, the internal-validation analysis of Section~\ref{sec:results-internal-validity} substantiates this, showing those contrasts persisting under resampling and structured perturbation.

Concretely, the empirical setting is a single English-language case study---ten ISO/IEC~27002 controls for AI4I4, from a text-only system description---with an expert-constructed gold standard and hallucination and canonicalization labels over the judged universe, and a fixed sample of stochastic LLM outputs under specific prompts, decoding settings, and model snapshots.
The question our post hoc analysis answers---whether fusing already-produced runs improves the order of review---is posed in the same terms for any system whose requirements are to be specified, and the fusion operators themselves are largely domain agnostic; what varies with different systems, domains, organizational contexts, and standards are the absolute precision--recall trade-offs and the magnitude (and relative ordering) of the ensembling gains, as judgments about relevance, atomicity, and merge/split decisions influence both the labels and the weighting signals derived from them.

Evaluation frames ensembling as a ranked-retrieval problem and uses IR-inspired measures at shifting operating points (precision, recall, and fall-out, plus combined metrics such as \(F_2\) and \(J_w\), and whole-ranking assessments such as AP and ROC AUC).
This metric-focused design is appropriate for comparing fusion operators, but it treats all valid requirements as equally important and does not measure downstream security efficacy, audit outcomes, prioritization value, or human review cost.
It is also, by construction, a closed-world evaluation: the evaluation is defined over the finite judged universe \(\mathcal{U}\) (Section~\ref{sec:supporting-dataset}), not the space of all conceivable requirements, and the true negatives are the judged hallucinated candidates that a ranking leaves unselected.
This is a construct-validity scope note more than a defect, and it is intrinsically mitigated: the headline quantities are relative contrasts---paired differences and normalized uplifts---computed within the same closed universe for every approach; that universe is fully adjudicated, as every candidate carries an expert label, so, unlike depth-limited retrieval pooling, it contains no unjudged items; and no approach can rank a candidate that lies outside it.
What remains genuinely universe relative is the absolute reading: recall and specificity are measured against \(\mathcal{U}\), valid requirements exist beyond it, and---since the space of invalid statements is effectively unbounded---an open-world notion of specificity would not be meaningful in any case.
That valid requirements exist beyond \(\mathcal{U}\) is not merely hypothetical: the exploratory source-study configurations excluded from this universe did surface further valid requirements, direct evidence that the requirement space extends past what any participating run produced.

The sampling design is also uneven across configurations: each model's base chain ran as four stochastic runs, whereas its cooler and warmer variants ran once each (Table~\ref{tab:configs}), so the base chain contributes more runs and could bias the comparison in its favor, since more draws yield both more votes in the fusion and more chances to surface rare valid requirements.
Two features of the design limit this effect without removing it.
First, configurations enter the comparisons through their per-run averages rather than individual runs, so a configuration's standing reflects its typical run rather than its luckiest draw or sheer number of draws.
Second, under Naive-Bayes fusion the vote of a model that behaves less consistently with the labeled corpus is discounted, so an over-sampled variant does not gain influence by volume alone unless its votes are also reliable; the weighting is per model, however, and does not equalize sampling among a model's own temperature variants.
Still, the more sampled variant retains a genuine recall advantage from its extra draws, so some temperature-driven differences may reflect the number of runs rather than the decoding settings alone; replications that equalize runs per variant---or subsampling-based sensitivity analyses---would further strengthen causal claims.

Relatedly, run-to-run variability is, in this design, less a threat to reproducibility than an exploited asset: because independent runs recover overlapping but not identical candidate sets, additional runs contribute complementary valid requirements and raise the recall available to fusion.
This complementarity is visible in the source study's consistency analysis~\cite{Yelmo2026AICyberReqGen}: within-configuration mean pairwise Jaccard similarity across runs ranges from 0.402 to 0.600 when hallucinations are included, rising to 0.647--0.788 once they are removed, while between-model similarity is substantially lower (\(\approx\)0.219, or 0.387 without hallucinations).
That is, runs converge more on true positives than on their idiosyncratic hallucinations (runs of the same model agree only moderately, and different models agree less still)---precisely the diversity, concentrated on the valid candidates, that voting-based fusion converts into ranked evidence.

The two fusion strategies differ in how much they depend on the data at hand.
Uniform fusion uses fixed, dataset-independent weights and requires no fitting; Naive-Bayes fusion derives its weights from the labeled corpus and is thus dataset dependent.
The validity of the latter rests on two controls reported in Section~\ref{sec:results-internal-validity}: out-of-bag cross-fitting, which trains the weights on a subset disjoint from the one used for evaluation, and a jackknife-like control-slice/run perturbation that applies the fixed full-corpus weights after removing one control slice and one run at a time.
Naive-Bayes is moreover proposed as a \emph{general} weighting rule---one that improves on Uniform fusion, which in turn improves on any single configuration---rather than as the best attainable fusion; better fitting weights exist for this corpus (Section~\ref{sec:implementation-strategies}), but pursuing them would amount to overfitting this particular dataset rather than improving the general rule.

Two of the reference curves against which the strategies are plotted---the pattern oracle and the numerical linear-fusion benchmark (Section~\ref{sec:evaluation-reference-curves})---might be mistaken for improved fusion approaches, but, like the other reference curves, they presuppose the observed labels and so are diagnostic bounds, not deployable methods.
The oracle marks an information bound that any vote-based scorer faces; the numerical benchmark estimates the headroom of linear weighting; and neither is a proposed method---in particular, the label-tuned benchmark contrasts with the article's proposed Naive-Bayes weighting, which is validated by cross-fitting.

A final caveat is temporal.
The absolute results and the model ranking are a snapshot of a fast-moving field, as the source study likewise noted.
We posit---and develop at greater length in Section~\ref{sec:impact-results}---that the durable part is the \emph{relative} advantage of ensembling over its own constituent runs, since that advantage is a property of run and model diversity rather than of any particular model, and should recur when today's constituents are replaced by newer ones.
This reading places our results within the broader evidence that aggregating the outputs of several imperfect LLMs improves on any single one~\cite{Wang2023SelfConsistency,Li2024MoreAgents,Chen2025LLMEnsembleSurvey}, a regularity not tied to a specific model generation; the magnitude of the gain for stronger future constituents remains an open empirical question.
Taken together, these constraints do not undermine the central contribution of this article---a controlled evaluation of voting and Naive-Bayes post hoc aggregation over LLM runs---but they motivate the replications and the risk-aware, interactive-workflow extensions outlined in Section~\ref{sec:conclusion}.

\section{Conclusion}
\label{sec:conclusion}

In this article, we have revisited a manually labeled corpus of cybersecurity requirements generated for ISO/IEC~27002 controls on the AI4I4 automotive logistics system, now focusing on a narrowly defined goal: how to improve the results by \emph{aggregating} multiple LLM outputs once they have already been produced.
By treating each LLM run as a black-box retrieval system over a canonicalized requirement universe, we framed post-generation ensembling as an information-retrieval data-fusion problem and evaluated voting- and quality-weighted ranking rules across 12 configurations from four model families.

The results bear out the perspective advanced in the introduction: treating the runs' disagreement as complementary evidence rather than evaluation noise, post hoc aggregation converts inter-run variability into a practical coverage advantage---and it does so chiefly by improving the \emph{order} in which requirements are reviewed, besides the sheer volume recovered.
Relative to the best single configuration, even the plainest form of agreement---Uniform fusion, which merely rewards candidate requirements supported by several runs---substantially increases recall at representative operating points without a matching rise in hallucinations.
Layering model-reliability-aware weights on top improves the ordering further: by discounting support from less reliable sources, Naive-Bayes fusion pushes valid requirements ahead of hallucinations more decisively---visible as higher AP and ROC AUC---and thus offers better practical precision--recall trade-offs, reaching comparable coverage at shallower review depths.
Internal validation indicates that these gains are stable, staying positive in at least 92\% of out-of-bag resamples and in every structured perturbation---though as an internal check it bears on robustness over this corpus, not external generality.

Taken together, the study supports a lightweight, post-generation workflow for AI-augmented security requirements generation---one that leaves the upstream models and prompts untouched and adds no model retraining or fine-tuning.
Rather than committing to a single carefully tuned configuration, practitioners can sample multiple stochastic runs across heterogeneous models and prompts and then apply voting-based fusion with a reliability-aware weighting layer (Naive-Bayes fusion with per-source noise rates) to order items for human review.
This preserves the diversity needed to surface rare or long-tail requirements while ranking likely valid content ahead of hallucinations.
Such a workflow also integrates naturally into established security requirements engineering processes: the fused ranking simply feeds their existing elicitation, prioritization, and inspection activities with a better ordered candidate queue.

Future work can extend this post-generation view in three directions.
First, replication across additional standards, system artifacts, and languages would test the generality of the ensembling findings.
Second, the reliability estimates could be complemented with control-specific checks that flag cross-model, plausible-but-off-scope requirements able to survive pure consensus.
Third, integrating these fusion operators into interactive analyst workflows would enable dynamic operating-point selection and clarify how ranked ensemble outputs translate into real compliance effort and risk reduction.
Beyond these research directions, a production deployment would naturally go further than the generic Naive-Bayes rule evaluated here, calibrating the fusion weights to its own models, corpus, and accumulating adjudication feedback.

In summary, this article shows that IR-inspired data fusion provides a principled and effective post hoc layer for aggregating LLM-generated cybersecurity requirements: the disagreement of stochastic generators is itself evidence, and it can be read after the fact, from the outputs alone.
Post-generation ensembling thus stands as an optimization layer in its own right---orthogonal to advances in prompting, fine-tuning, retrieval augmentation, or larger foundation models---ready to complement whichever of them a pipeline adopts.

\appendix
\section{Reproducible Research}
All artifacts required to replicate the evaluation reported here---the frozen candidate corpus, the ensembling code, and the computed metrics, curves, and result tables---are archived in a dedicated Zenodo deposit for this study (\url{https://doi.org/10.5281/zenodo.21496481}).
This deposit is separate from the reproducible-research package of the source study that produced the underlying LLM runs and gold standard~\cite{Yelmo2025RRPackage}, which we reuse with its authors' subsequent annotation corrections.
Our artifacts refer to each candidate requirement---the 72 vetted valid requirements and the 111 hallucinated ones that together make up the judged universe---only by an identifier; the corresponding requirement text is resolved against that source package.

All ensembling reported here is offline post-processing over that frozen corpus: no new LLM calls are made and no requirements are regenerated.
Every strategy operates only on the recorded per-run presence of each candidate, so the archived data fully determine the reported results.

Because the underlying repository is a broader benchmarking toolkit, it computes more than this article reports---further ranking metrics, additional weighting families, and numerical optimizers among them.
The article draws only on Uniform and Naive-Bayes fusion evaluated with the core metrics; the repository's README file (supplied in the Zenodo record) documents the remaining strategies, metrics, and analyses, along with the mapping from their internal names to the labels used here, as well as a complete directory map, setup instructions, and usage examples.

\section*{Acknowledgment}
\label{sec:ai-usage-disclosure}
\textbf{AI usage disclosure.}
During the preparation of this article, the authors worked in an iterative cycle with AI assistants---Anthropic Claude Code, OpenAI ChatGPT, GitHub Copilot, and OpenAI Codex---on writing, coding, and research tasks spanning all sections of the manuscript: the text and the analysis code grew through continuously interleaved author- and AI-written revisions, each contribution reviewed and built upon in the next round, with dedicated passes by adversarially instructed AI agents complementing the authors' own reviews.
All article contents and all significant code were revised by the authors, who are fully accountable for them.

\textbf{Source dataset.}
This study builds on the dataset of the authors' previous work~\cite{Yelmo2026AICyberReqGen}: the recorded LLM runs and the expert-adjudicated gold standard are reused here, subject only to the annotation corrections recorded since.

\phantomsection
\label{sec:references}
\bibliographystyle{IEEEtran}
\bibliography{references}

\label{sec:biographies}
\begin{IEEEbiography}[{\includegraphics[width=1in,height=1.25in,clip,keepaspectratio]{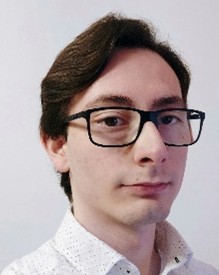}}]
{SANTIAGO PEREZ-ACUNA}~is currently a Predoctoral Researcher with the Department of Telematics Engineering, Universidad Politécnica de Madrid (DIT-UPM), Spain. He received the B.Sc. degrees in computer engineering and in business administration from the University of Vigo, Spain, and the M.Sc. degree in network and telematic services engineering from the Universidad Politécnica de Madrid. His research focuses on the application of artificial intelligence to cybersecurity in industrial software systems. His main interests include AI-assisted requirements engineering, large language models, and the development of secure and trustworthy software in Industry 4.0 contexts.
\end{IEEEbiography}

\begin{IEEEbiography}[{\includegraphics[width=1in,height=1.25in,clip,keepaspectratio]{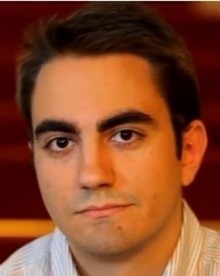}}]
{YOD-SAMUEL MARTÍN}~(M'14)~received the M.Sc. degree in telecommunications engineering from the Universidad Politécnica de Madrid (UPM), Spain, in 2004, the M.Phil. degree in 2006, and the Ph.D. degree in 2022.

He has held different research and lecturing positions at UPM, in the Center for Open Middleware (COM), Information Processing and Telecommunications Center (IPTC), and Departamento de Ingeniería de Sistemas Telemáticos (DIT), where he is currently an Assistant Professor. His research interests include the management of different categories of extra-functional requirements (cybersecurity, privacy, accessibility, usability, internationalization) throughout the development lifecycle.
\end{IEEEbiography}

\begin{IEEEbiography}[{\includegraphics[width=1in,height=1.25in,clip,keepaspectratio]{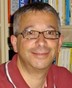}}]
{JUAN C. YELMO}~received the M.S. and Ph.D. degrees in telecommunications engineering from the Universidad Politécnica de Madrid, Madrid, Spain, in 1991 and 1996, respectively. He is currently an Associate Professor with the Universidad Politécnica de Madrid, where he has over 25 years of experience in software and internet services engineering. He has participated in numerous national and international research projects and has been involved in technology transfer and innovation activities, including research contracts with international companies, standardization efforts, open innovation initiatives, and consulting for university spin-offs. His current research interests include cybersecurity and AI-augmented software engineering.
\end{IEEEbiography}

\EOD

\end{document}